\documentclass[aps,prd,twocolumn,superscriptaddress,amsmath,showpacs]{revtex4-1}
\usepackage{graphicx}
\usepackage{dcolumn}
\usepackage{bm}
\usepackage{natbib}

\newcommand{\be}{\begin{equation}}
\newcommand{\ee}{\end{equation}}
\newcommand{\bea}{\begin{eqnarray}}
\newcommand{\eea}{\end{eqnarray}}

\newcommand{\Mpc}{{\rm ~Mpc}}

\urldef\mgcamb\url{http://www.sfu.ca/~aha25/MGCAMB.html}

\begin{document}
\title{Is PLANCK consistent with primordial deuterium measurements ?}
\author{Laura Salvati}
\affiliation{Physics Department and INFN, Universit\`a di Roma ``La Sapienza'', Ple Aldo Moro 2, 00185, Rome, Italy}

\author{Najla Said}
\affiliation{Physics Department and INFN, Universit\`a di Roma ``La Sapienza'', Ple Aldo Moro 2, 00185, Rome, Italy}

\author{Alessandro Melchiorri}
\affiliation{Physics Department and INFN, Universit\`a di Roma ``La Sapienza'', Ple Aldo Moro 2, 00185, Rome, Italy}

\begin{abstract}
The recent measurements of the Cosmic Microwave Background Anisotropies provided by the Planck
satellite experiment have significantly improved the constraints on several cosmological
parameters. In this brief paper we point out a small but interesting tension present between recent 
values of the  primordial deuterium measured from quasar absorption line systems and the same value inferred, albeit indirectly, from the Planck measurements assuming $\Lambda$CDM and Big Bang Nucleosynthesis. 
Here we discuss this tension in detail investigating the possible new physics that 
could be responsible for the tension. We found that, among $8$ extra parameters, only an anomalous lensing component and a closed universe could change the Planck constraint towards a better 
consistency with direct deuterium measurements.
\end{abstract}


\maketitle

\section{Introduction} \label {sec:intro}

The standard Big Bang Nucleosynthesis (BBN, hereafter) relates the abundance of the light elements created in the early Universe with key cosmological parameters such as the baryon energy density $\Omega_b$ or the number of relativistic species $N_{\rm eff}$. Given a set of cosmological parameters, BBN can theoretically predict the primordial abundances with great accuracy and these predictions can then be compared to the observational measurements for an overall test of the theory.

The primordial abundance of light elements is generally measured with astrophysical observations (see e.g. ~\cite{Iocco1} and references therein), by looking at regions in the Universe free from contamination from stellar nucleosynthesis or taking in account the depletion or accretion of the elements abundances due to these processes. In the last years several efforts were made in order to achieve a robust treatment of systematics, in order to obtain stronger and reliable constraints on these measurements~\cite{Aver1}.\\
Most of the focus is on the primordial values of $^4He$, $D$ and $^7Li$, and, in general, there is a good agreement between the values of the cosmological parameters inferred from each measurement, with the exception of $^7Li$ (there is, indeed, a tension for what concerns the value of $^7Li$ that needs still to be explained, see e.g. ~\cite{Fields, Iocco2}).

Recently,~\cite{Pettini-Cooke} measured the $D$ abundance with impressive accuracy from a metal poor damped Lyman alpha system (DLA), namely the QSO SDSS J1419+0892. In this work, the following value for deuterium abundance is given: $\log(D/H)=−4.596\pm0.008\pm0.003$ (statistical and systematic 1$\sigma$ errors). Summing in quadrature the errors, as also done in~\cite{Pettini-Cooke}, one obtains for the D abundance a $68 \%$ c.l. constraint of $(D/H)=2.535\pm0.050\cdot 10^{-5}$. Previous measurements of the $D$ abundance are listed in~\cite{Iocco1}, that derives a mean value from them of $(D/H)=2.87^{+0.22}_{-0.21}\cdot 10^{-5}$, showing an error that is almost 5 times larger than that of Pettini $\&$ Cooke.

Another indirect way to accurately measure the primordial elements is based on the determination
of the baryon density through measurements of the Cosmic Microwave Background anisotropy angular power spectrum and to then derive the primordial abundances by assuming standard BBN.
The recent measurements made by the Planck satellite mission sampled the CMB power spectrum with very high accuracy up to very small angular scales, at multipole $\ell=2500$~\cite{PlanckXVI} and provided new, more stringent, bounds on the baryon density, with $\Omega_b h^2=0.02205\pm0.00028$. 
This value, under the assumption of standard BBN, as explained below, brings to a constraint on the primordial Deuterium abundance of $(D/H)=2.661^{+0.06}_{-0.05}\cdot 10^{-5}$ (see \cite{PlanckXVI}), with a central value that is more than two standard deviations away from the value reported in \cite{Pettini-Cooke}.

The tension is therefore not highly statistically significant, taking into account small systematics that could clearly be present both in the CMB and Deuterium measurements.
Moreover, as pointed out in \cite{eleo014}, the experimental value of the radiative capture reaction
$d(p,\gamma)^3He$, used in BBN codes as  \texttt{PArthENoPE} code~\cite{parthenope} and that converts deuterium into helium, is in disagreement with ab initio theoretical calculations, based on models for the nuclear electromagnetic current derived from realistic interactions.
Systematic may therefore be also present in fits of present experimental data in the BBN 
energy range ($10$ - $300$ keV).

However, it is certainly timely and interesting to investigate if any deviation from the theoretical assumed model could better accommodate the two 
determinations, i.e. to consider the tension as a possible hint for new physics and assuming 
that systematics are under control in current experimental measurements.

The aim of this paper is to study in detail the correlations between the primordial abundance of
Deuterium, derived by the Planck data under the assumption of standard BBN, and other, non standard, cosmological parameters. We will therefore first identify the non-standard parameters that could
bring the Planck $D$ value in more agreement with the value reported by \cite{Pettini-Cooke}.
The subsequent, natural, step will be to include the direct $D$ measurement in the analysis and
to derive constraints on the non-standard parameters for combined Planck+$D$ measurements.

Furthermore, Planck results show some discrepancy with previous experiments as WMAP~\cite{wmap}, ACT~\cite{act}, SPT~\cite{spt} on the values of key cosmological parameters (i.e. matter density, lensing amplitude or number of relativistic species in early Universe), as studied in~\cite{DiValentino, Said}. Therefore it's interesting to explore whether a change in the $D$ abundance can help to solve this tensions.

The paper is organized as follows: in Section 1 we describe the data analysis method, in Section 2 we report the results of our analysis, while in Section 3 we draw our conclusions.

\section{Data analysis method}\label{sec:data}

Our main CMB dataset consists in the Planck public data release of March 2013 \cite{PlanckXV}. We use the combination of Planck temperature data plus WMAP9 polarization at low$\ell$ (Planck+WP) to evaluate the abundances of light nuclei produced during the Big Bang Nucleosynthesis, in particular $D$. These abundances are obtained using the \texttt{PArthENoPE} code~\cite{parthenope}, as a function of the baryon density $\omega_b=\Omega _bh^2$ and the number of relativistic species $N_{\rm eff}$. This code uses values for nuclear reaction rates, fundamental constants and particle masses that leads to a theoretical error for the obtained light elements abundances. The obtained values are compared with the $D$ abundance measured by direct astrophysical observations made by Pettini $\&$ Cooke~\cite{Pettini-Cooke}.\\
For the analysis method we use the publicly available Monte Carlo Markov Chain package \texttt{cosmomc} \cite{Lewis:2002ah} which relies on a convergence diagnostic based on the Gelman and Rubin statistic. We use the latest version (March 2013) which includes the support for the Planck Likelihood Code v1.0 (see \url{http://cosmologist.info/cosmomc/}) and implements an efficient sampling of the space using the fast/slow parameters decorrelation \cite{Lewis:2013hha}.
We first consider the $\Lambda$CDM model varying its six parameters: the baryon and cold dark matter densities $\Omega_{ b}h^2$ and $\Omega_{ c}h^2$, the ratio of the sound horizon 
to the angular diameter distance at decoupling $\theta$, the re-ionization optical depth $\tau$, the scalar spectral index $n_S$, and the overall normalization of the spectrum $A_S$ at $k=0.05\Mpc^{-1}$. We consider purely adiabatic initial conditions and
we impose spatial flatness. We fix also the number of neutrinos, assumed massless, at the standard value $N_{\rm eff}=3.046$~\cite{Mangano}.\\
Subsequently we study several extensions to the standard model, considering $8$ new parameters: the amplitude of the lensing power $A_{\rm L}$~\cite{Calabrese}, the equation of state parameter for the dark energy $w$, the ratio between tensor and scalar perturbations $r$, the running of the spectral index $n_{\rm r}=d\,n_S/d\ln k$, the curvature parameter $\Omega _{\rm k}$, the mass and number of neutrinos $m_{\nu}$ and $N_{\rm eff}$ and the parameter for iso-curvature perturbations $\alpha _1$. We consider $13$ theoretical frameworks by combining different choices of parameters.\\

\section{Results}\label{sec:results}

\subsection{CMB data only}
We first consider the $\Lambda$CDM model to evaluate the $D$ abundance from the CMB temperature anisotropies. By making use of the \texttt{PArthENoPE} BBN code, one can obtain the value of $D$ in function of $\Omega_bh^2$ and $N_{\rm eff}$. 
By using $N_{\rm eff}=3.046$ and the result for $\Omega_bh^2$ from Planck we obtain a value of $D$: $(D/H)=2.66^{+0.06}_{-0.05}\cdot 10^{-5}$.

Comparing this value with the one obtained by direct observations of Pettini $\&$ Cooke, $(D/H)=2.535\pm0.050\cdot 10^{-5}$, we get a tension at $1\div2 \,\sigma$. As discussed in the previous section, in order to further investigate this tension, we consider $13$ new models, extensions of the standard one, and determine the new distributions for $(D/H)$.
In Table \ref{table:1} are reported the values of $(D/H)$ for each case considered in our analysis. 

We can summarize our results by identifying three different classes:

\begin{enumerate}
 \item in models where $w$, $r$, $n_{\rm r}$, $\alpha_1$ are free to vary there is no variation (or at least a very small variation) in the Planck values of Deuterium with respect to the standard model. These extensions of the $\Lambda$CDM model, indeed, do not affect significantly the $\Omega_bh^2$ value, resulting in a negligible change of the $D$ predicted value (see Table \ref{table:1}). 
 
 \item in models with $A_{\rm L}$ or $\Omega _k$ as free parameters we see a strong variation with respect to the standard model and we obtain values of Deuterium that are now in good agreement, even within $1\,\sigma$, with the value of Pettini $\&$ Cooke;
 
 \item in the two models in which $\sum m_{\nu}$ and $N_{\rm eff}$ are free to vary the new Deuterium distributions are even further away from the one obtained by Pettini $\&$ Cooke with respect to the standard model. Clearly, a Planck+$D$ analysis does not prefer a non-zero neutrino mass
 or a non-standard value for $N_{\rm eff}$.
 \end{enumerate} 

\begin{center}
\begin{figure*}[h!] 
 \includegraphics[scale=0.35]{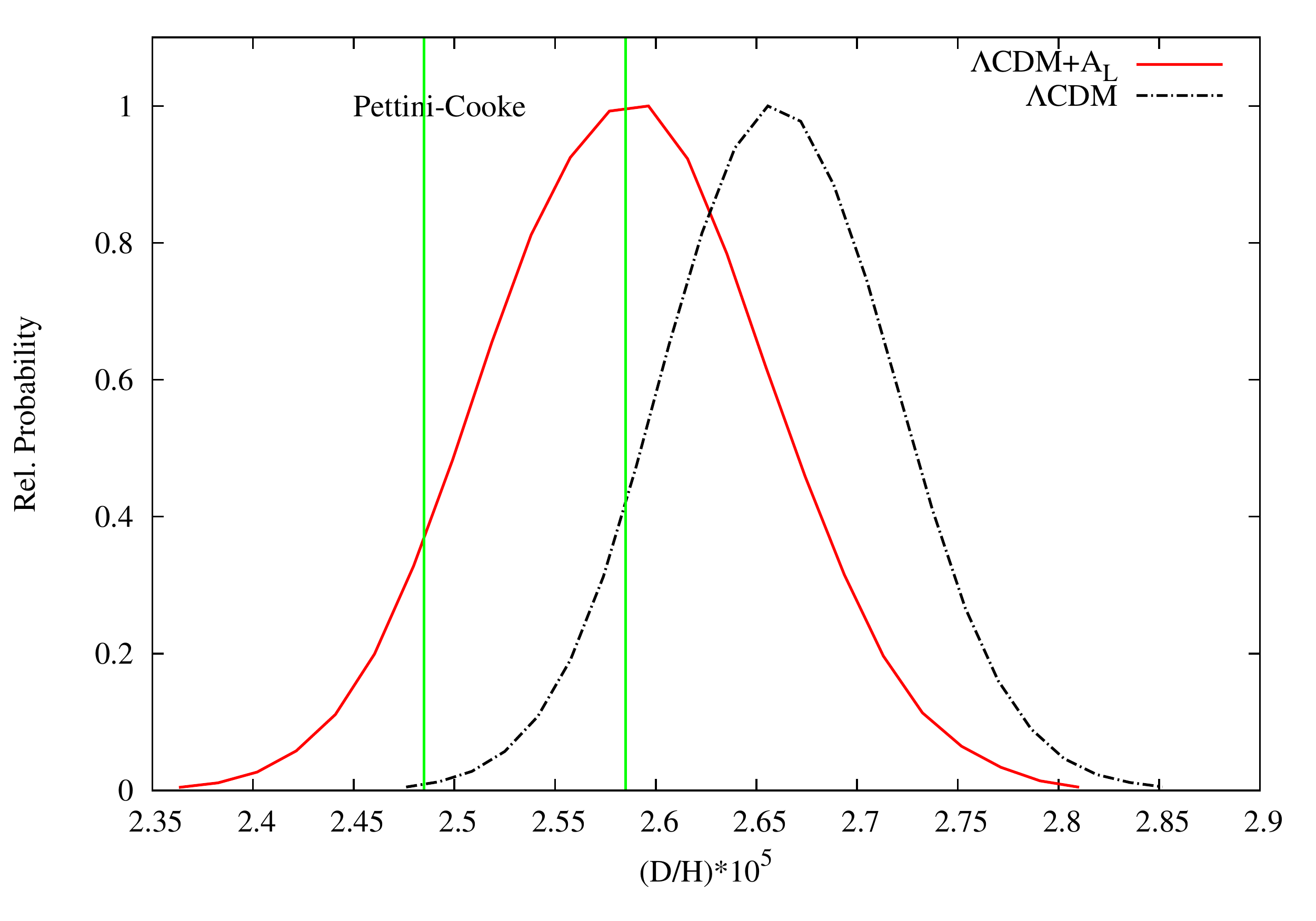} 
 \includegraphics[scale=0.35]{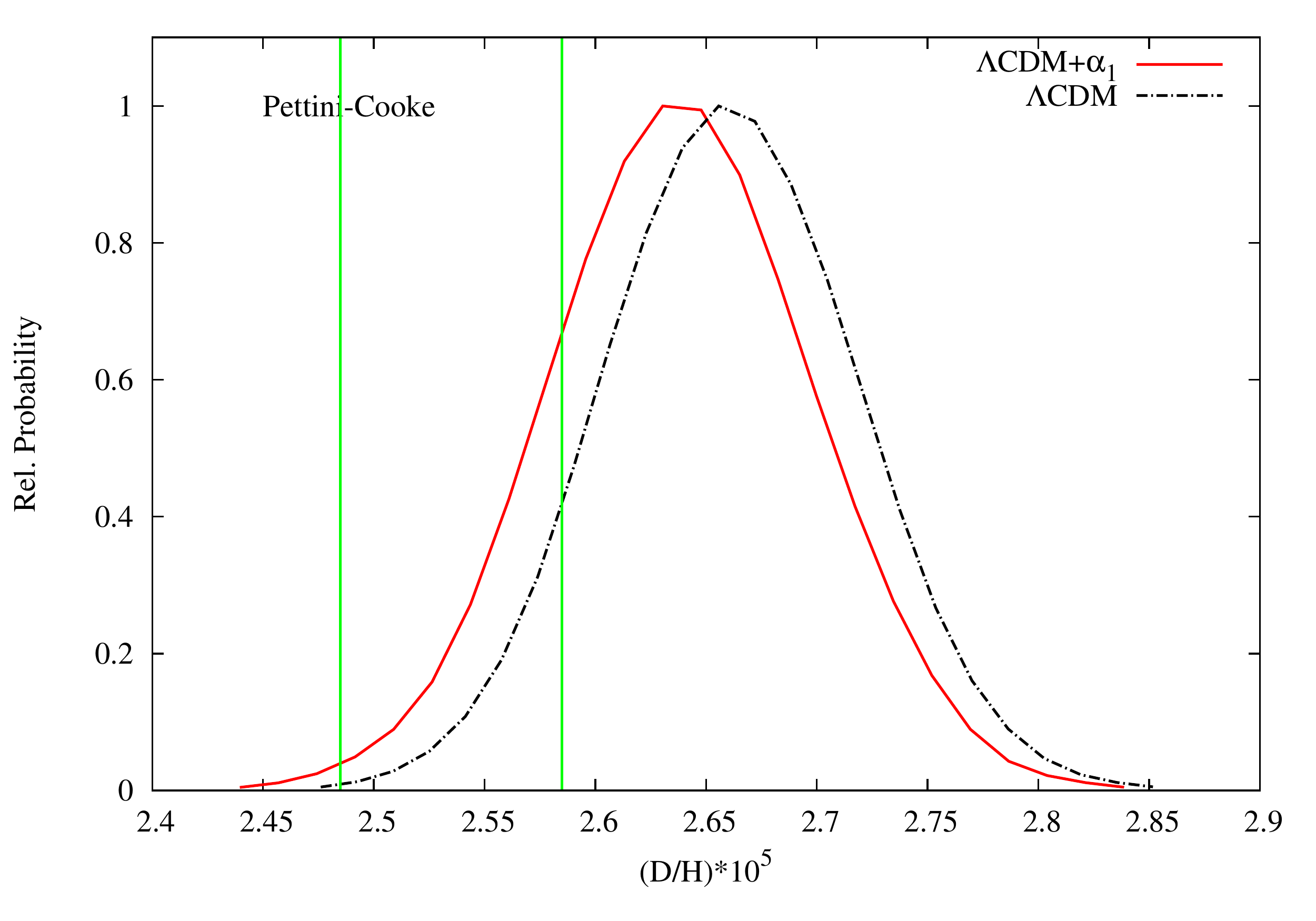} \\
 \includegraphics[scale=0.35]{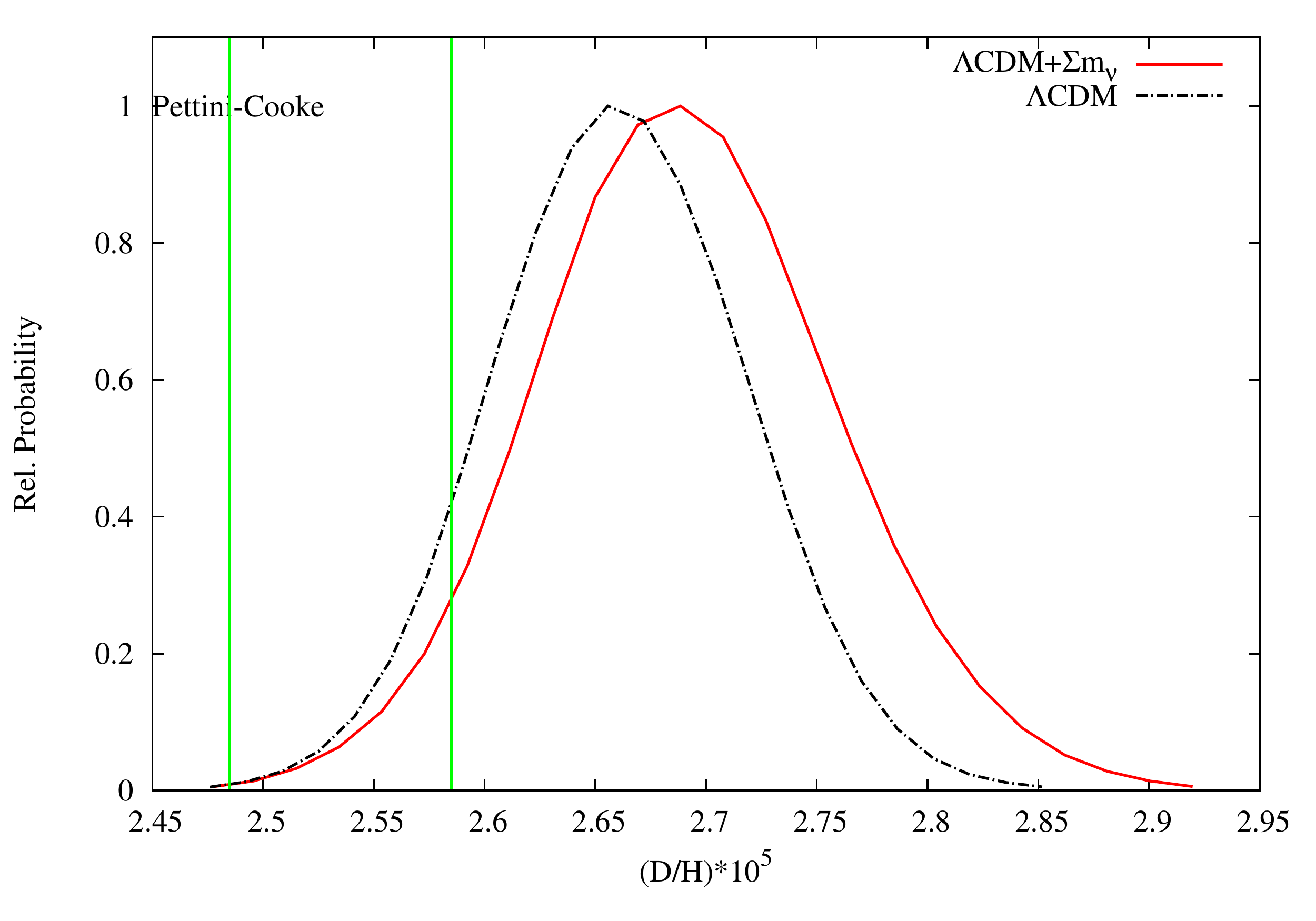} 
 \includegraphics[scale=0.35]{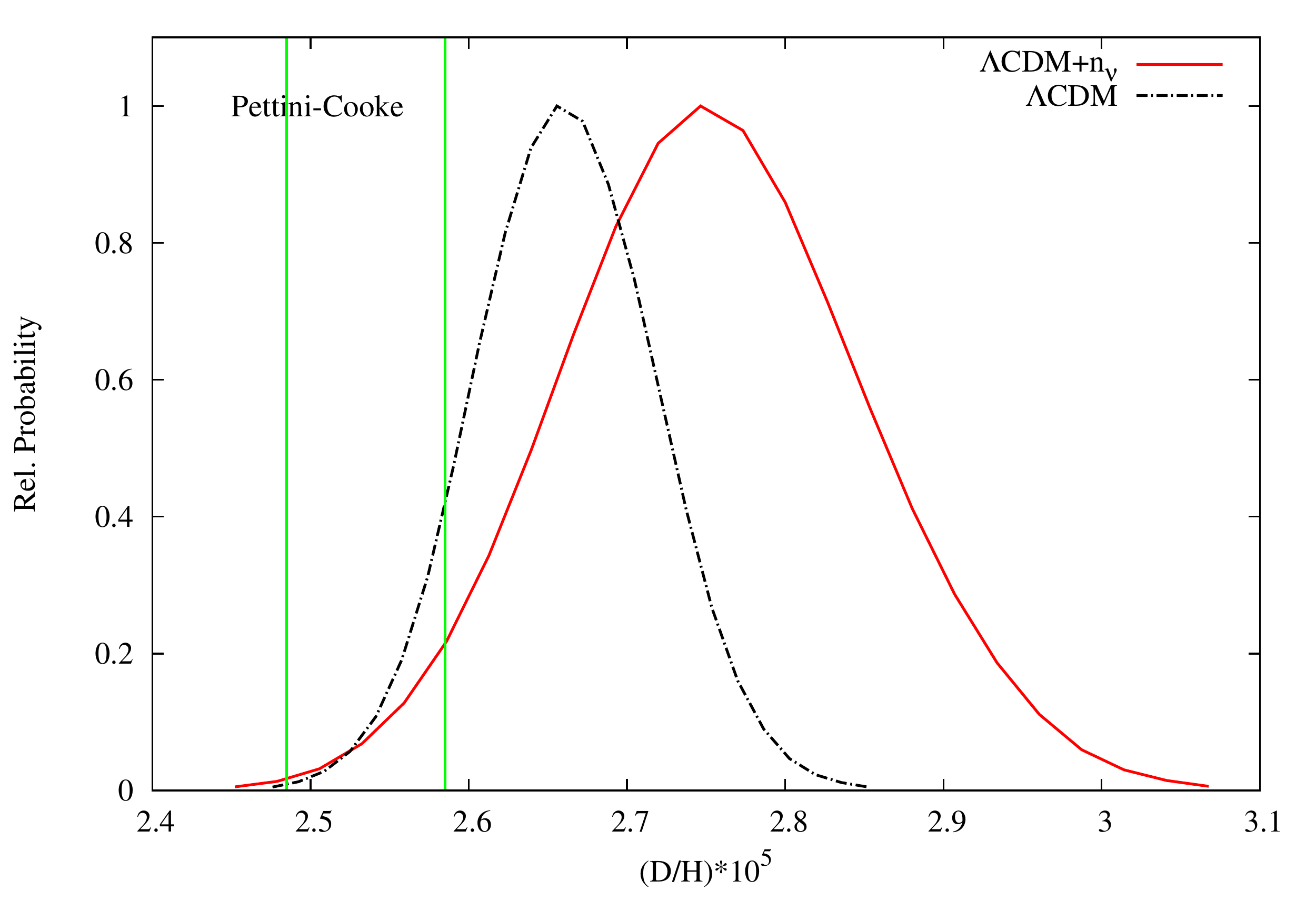} 
 \caption{\footnotesize{In this figure we report the posterior distributions for Deuterium in different models (red line): $\Lambda$CDM$+A_{\rm L}$, $\Lambda$CDM$+\alpha_1$, $\Lambda$CDM$+m_{\nu}$ and $\Lambda$CDM$+n_{\nu}$, always comparing the results with the ones obtained in the standard $\Lambda$CDM model (black line). We report also the $1\sigma$ limits for the results found by Pettini $\&$ Cooke (green lines).}}
 \label{fig:ap1}
\end{figure*}
\end{center}

\begin{center}
\begin{figure*}[h!]
 \includegraphics[scale=0.35]{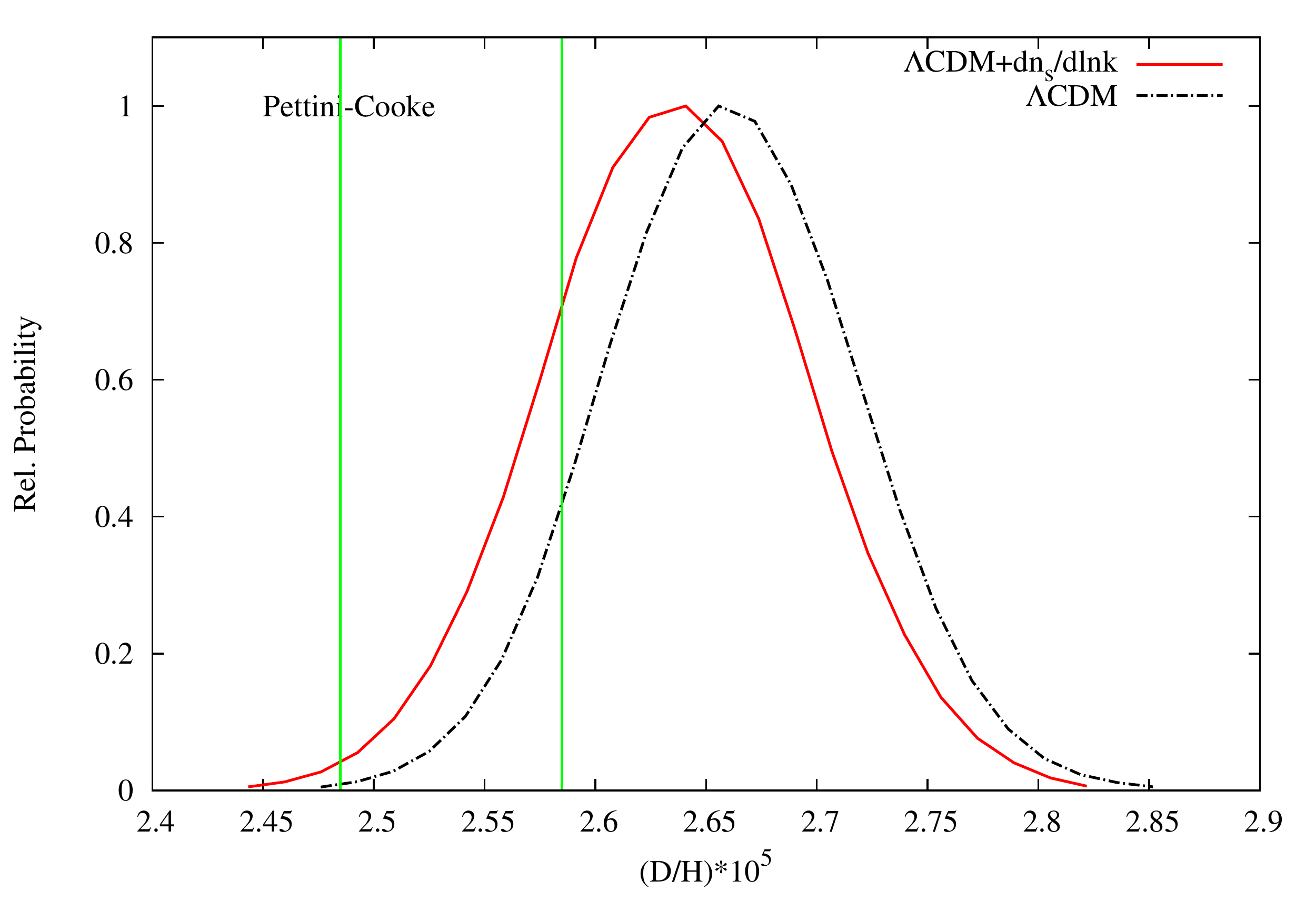} 
 \includegraphics[scale=0.35]{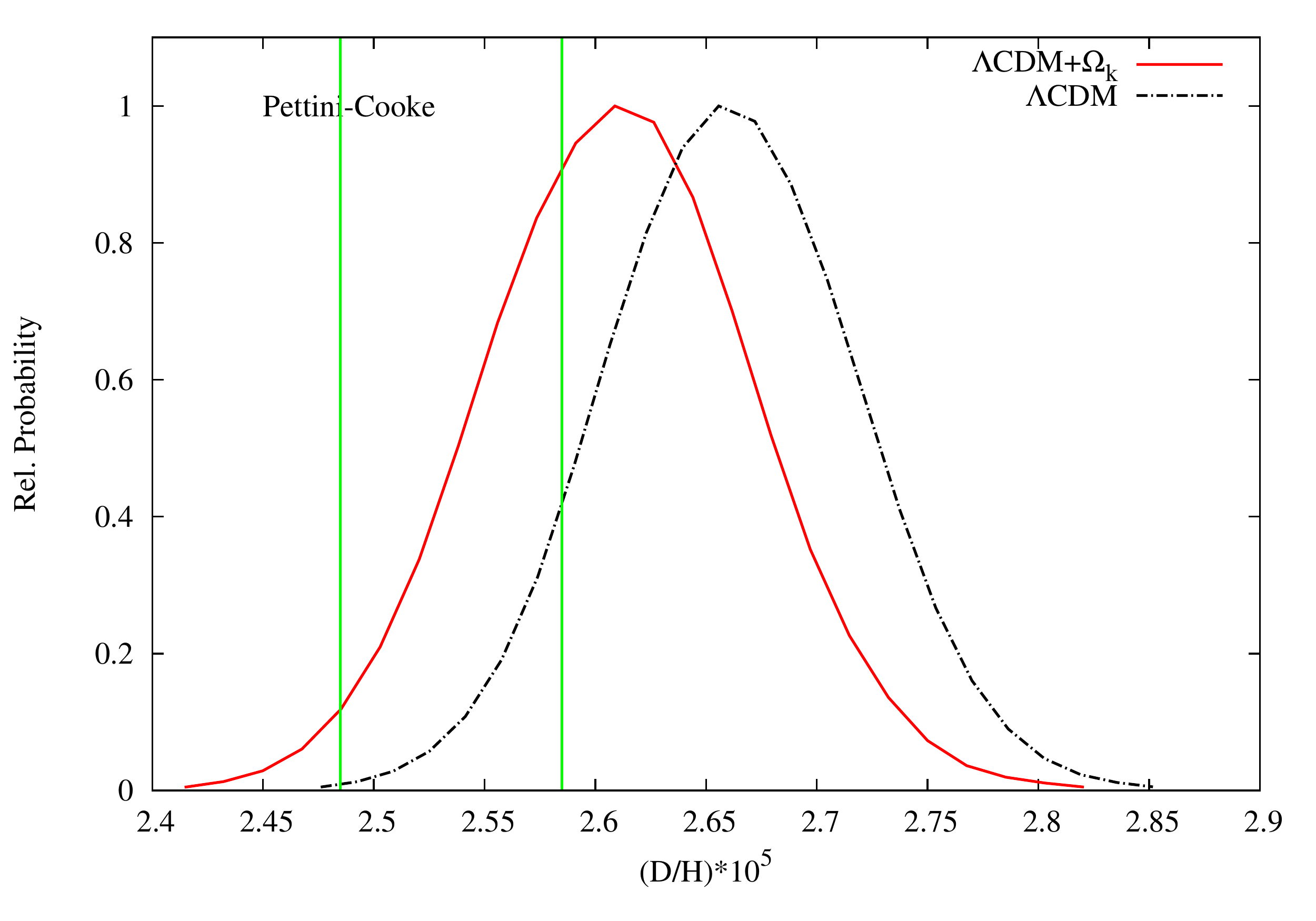} \\
 \includegraphics[scale=0.35]{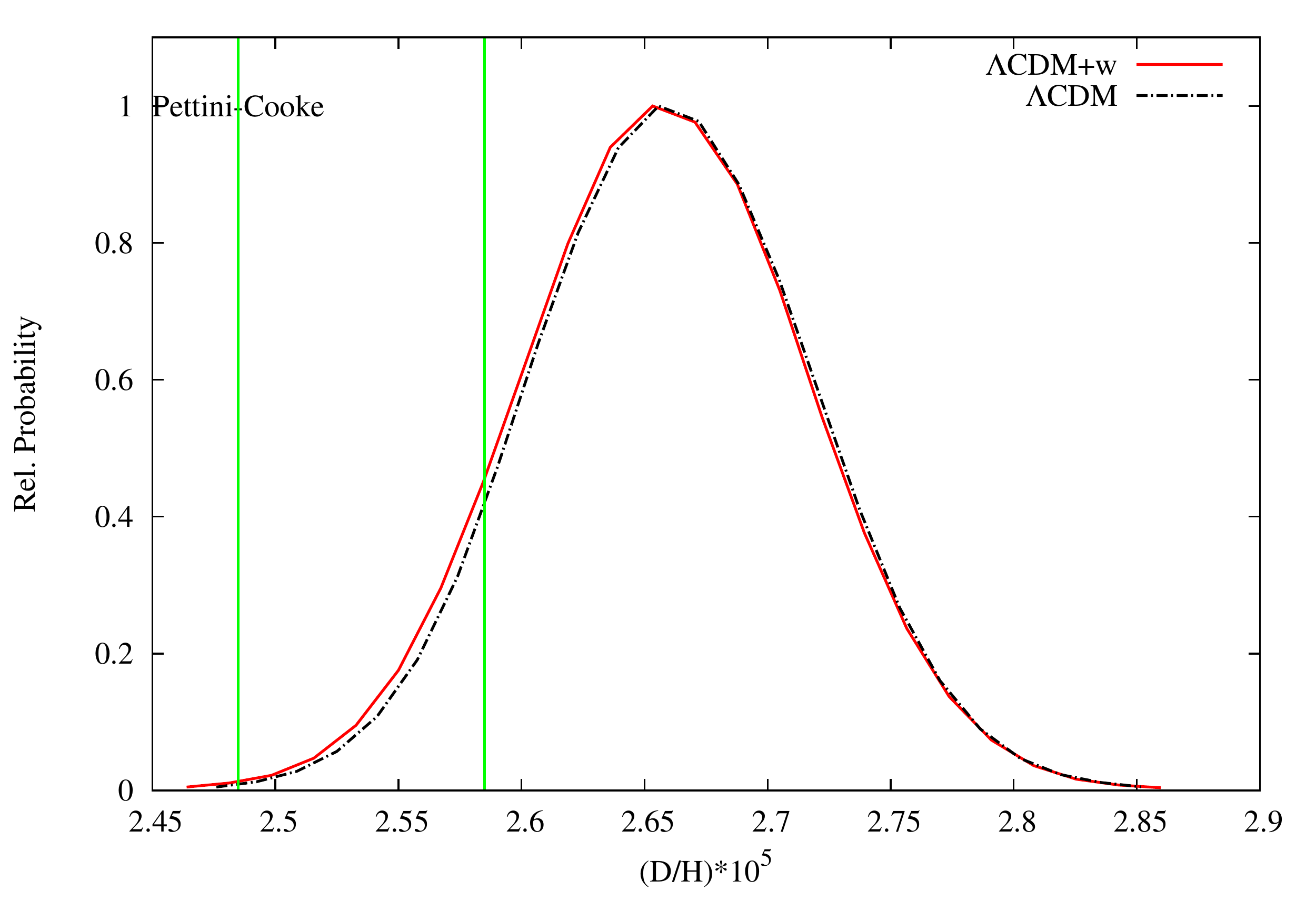} 
 \includegraphics[scale=0.35]{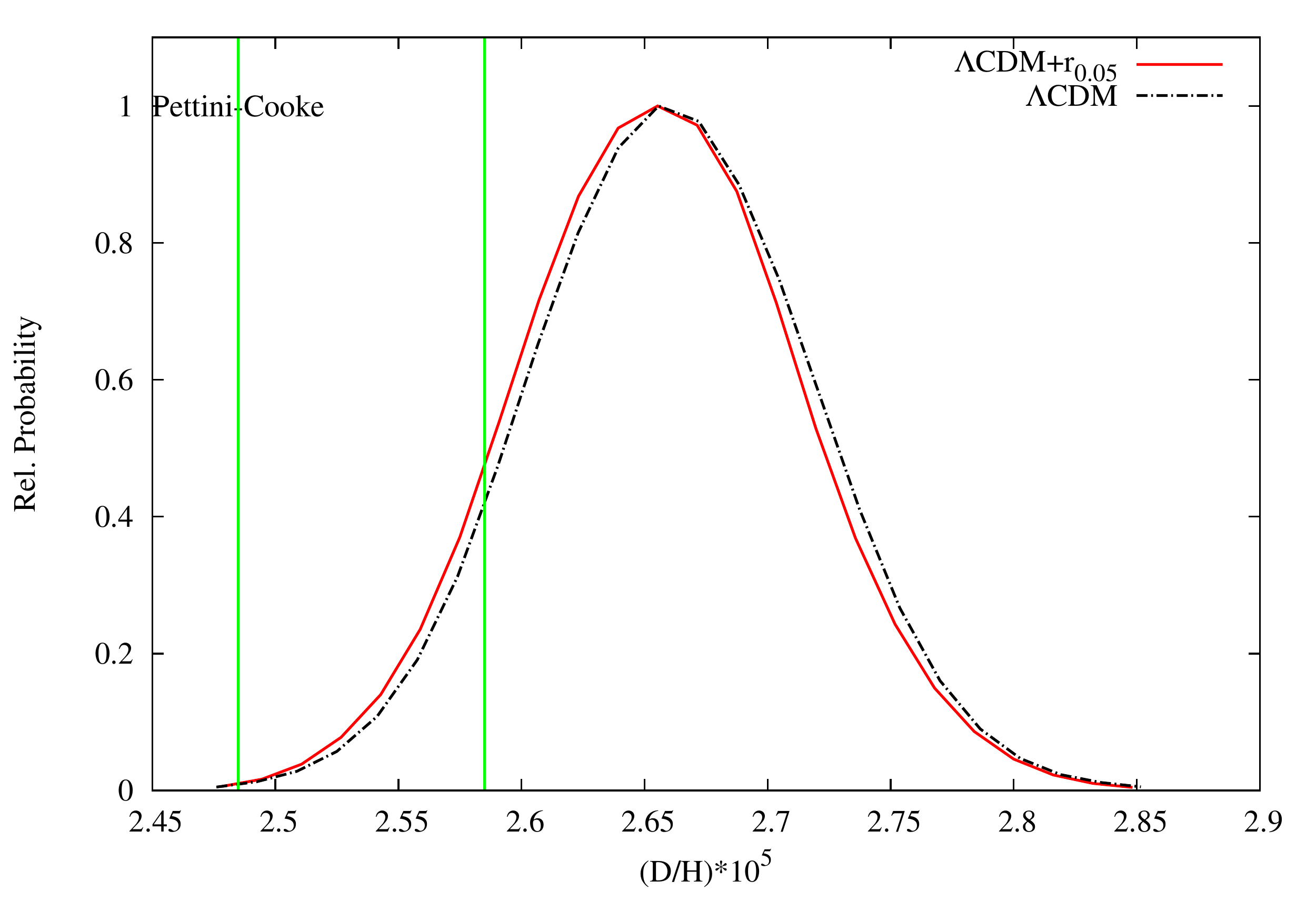} 
 \caption{\footnotesize{Here are reported the posterior distributions for deuterium, considering different models (red line): $\Lambda$CDM$+n_{\rm r}$, $\Lambda$CDM$+\Omega_{\rm k}$, $\Lambda$CDM$+w$ and $\Lambda$CDM$+r$, still comparing the results with the standard model (black line). We report also the $1\sigma$ limits for the results found by Pettini $\&$ Cooke (green lines).}}
 \label{fig:ap2}
\end{figure*}
\end{center}

\begin{center}
\begin{figure*}[h!]
 \includegraphics[scale=0.35]{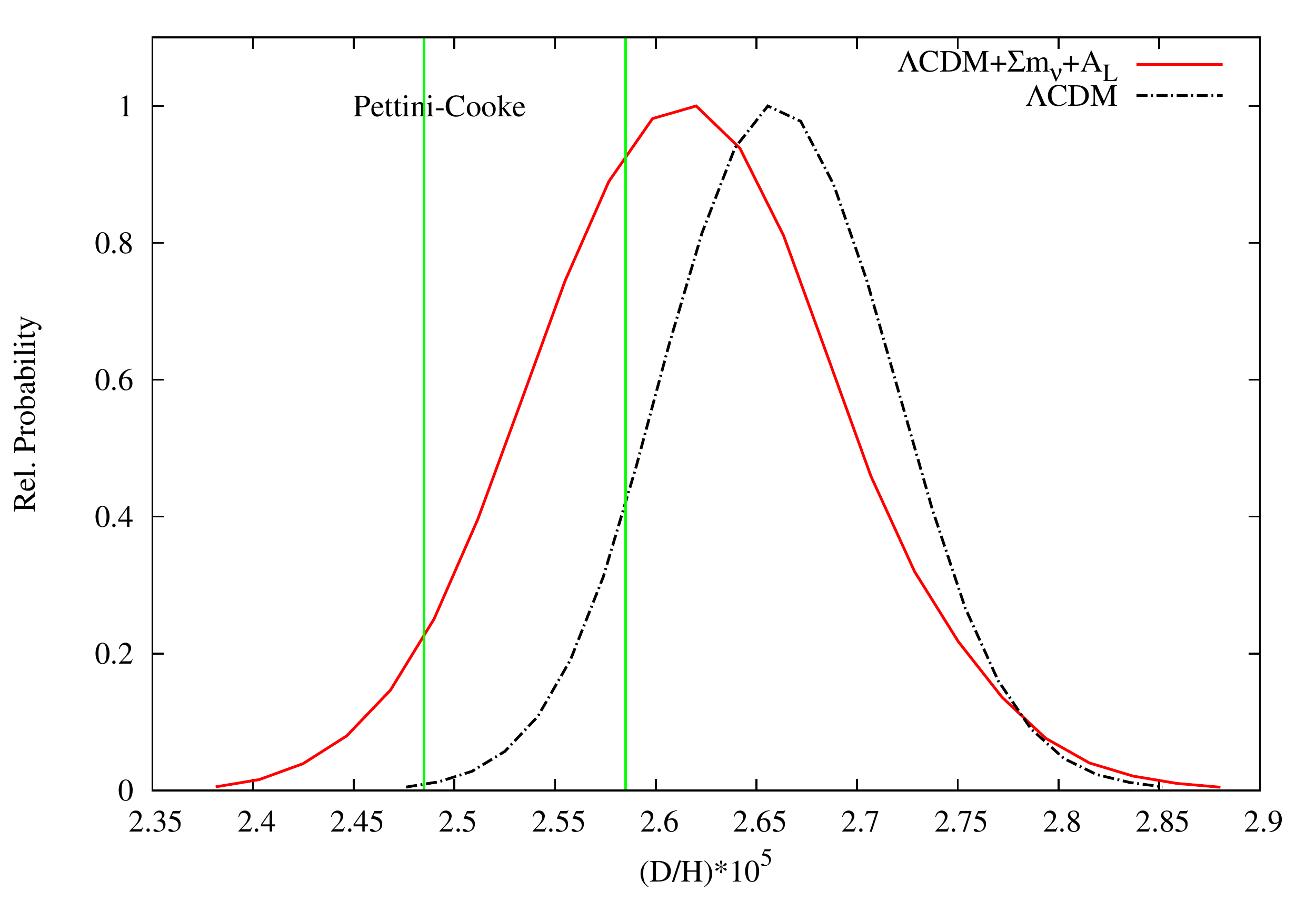} 
 \includegraphics[scale=0.35]{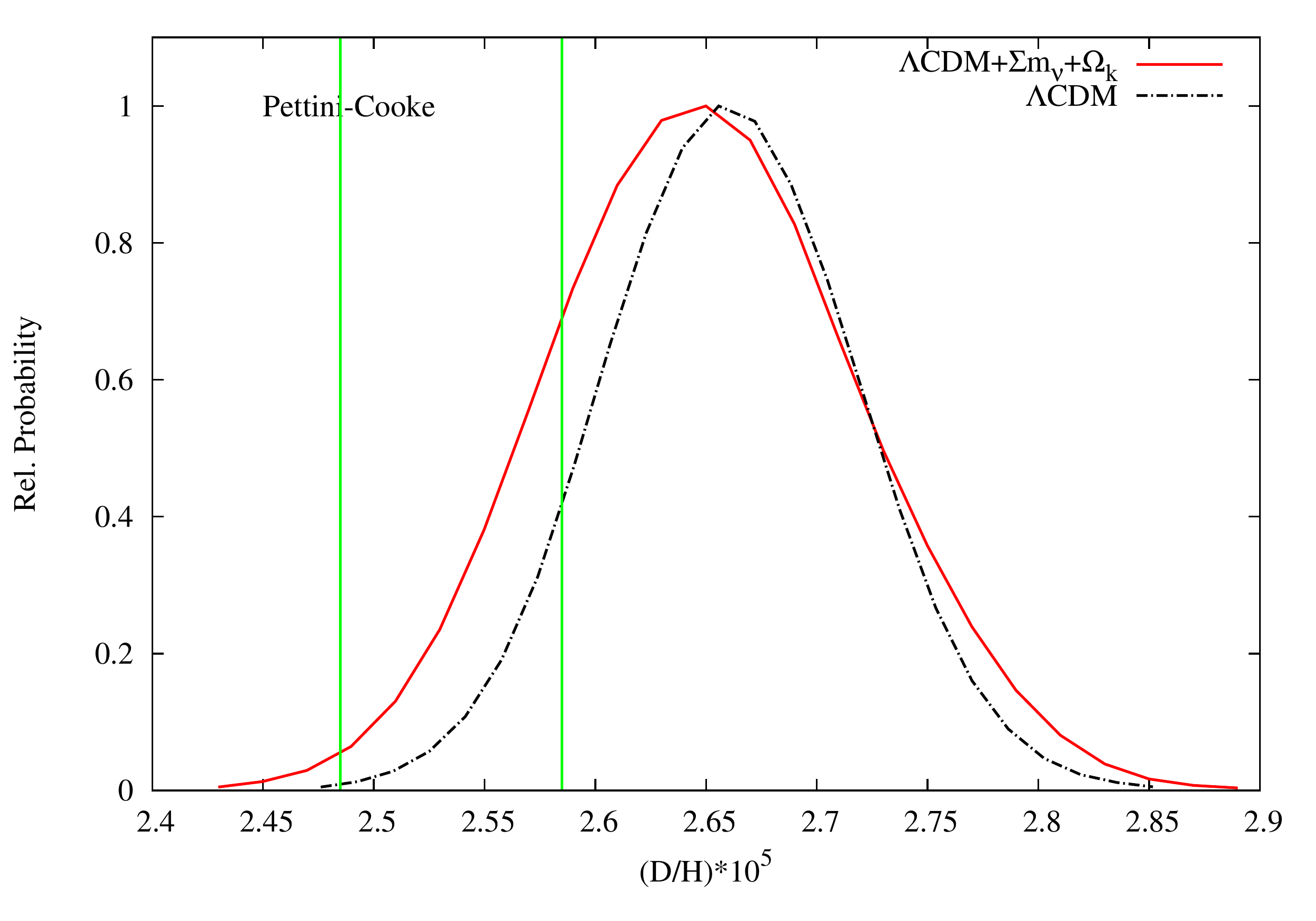} \\
 \includegraphics[scale=0.35]{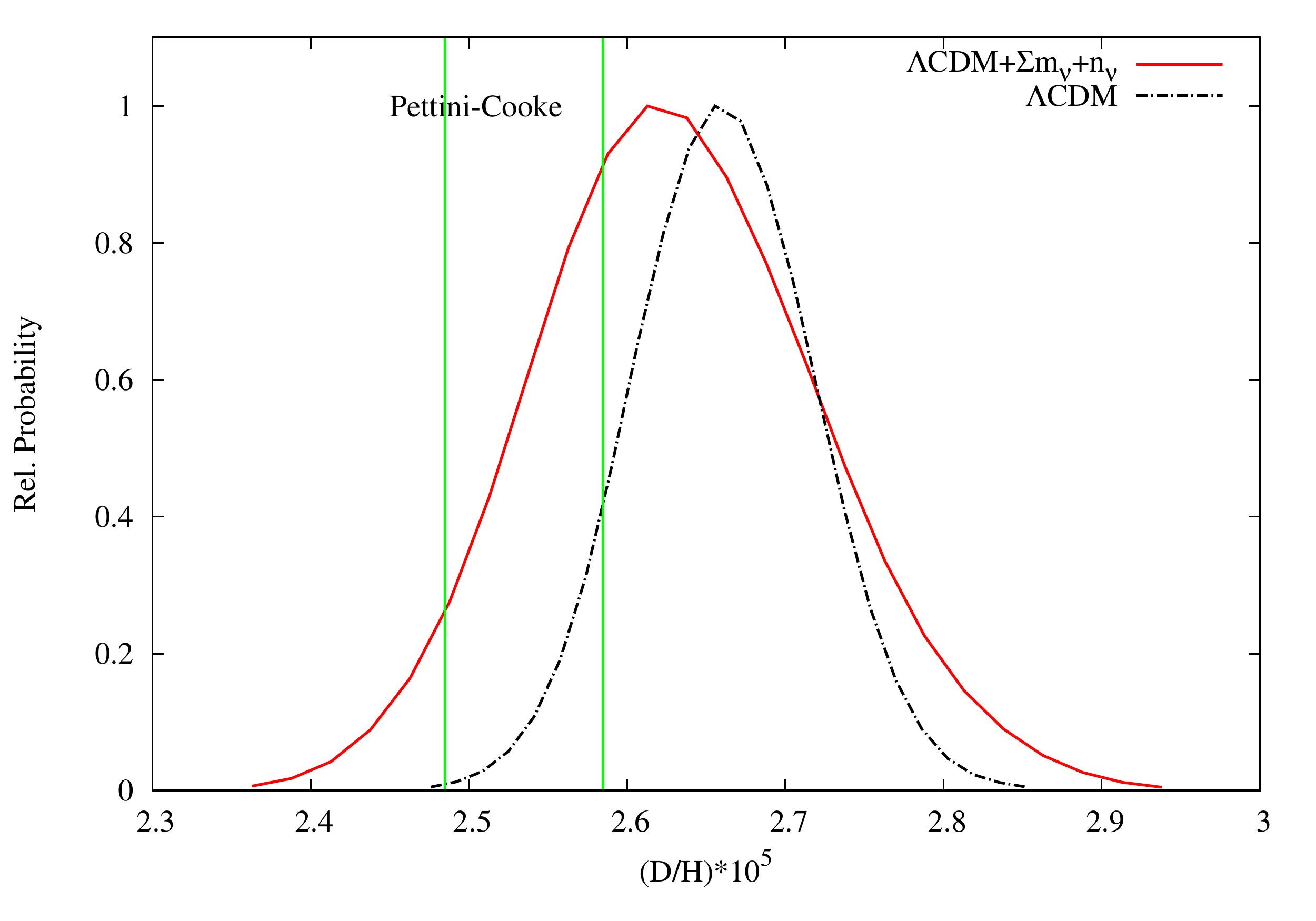} 
 \includegraphics[scale=0.35]{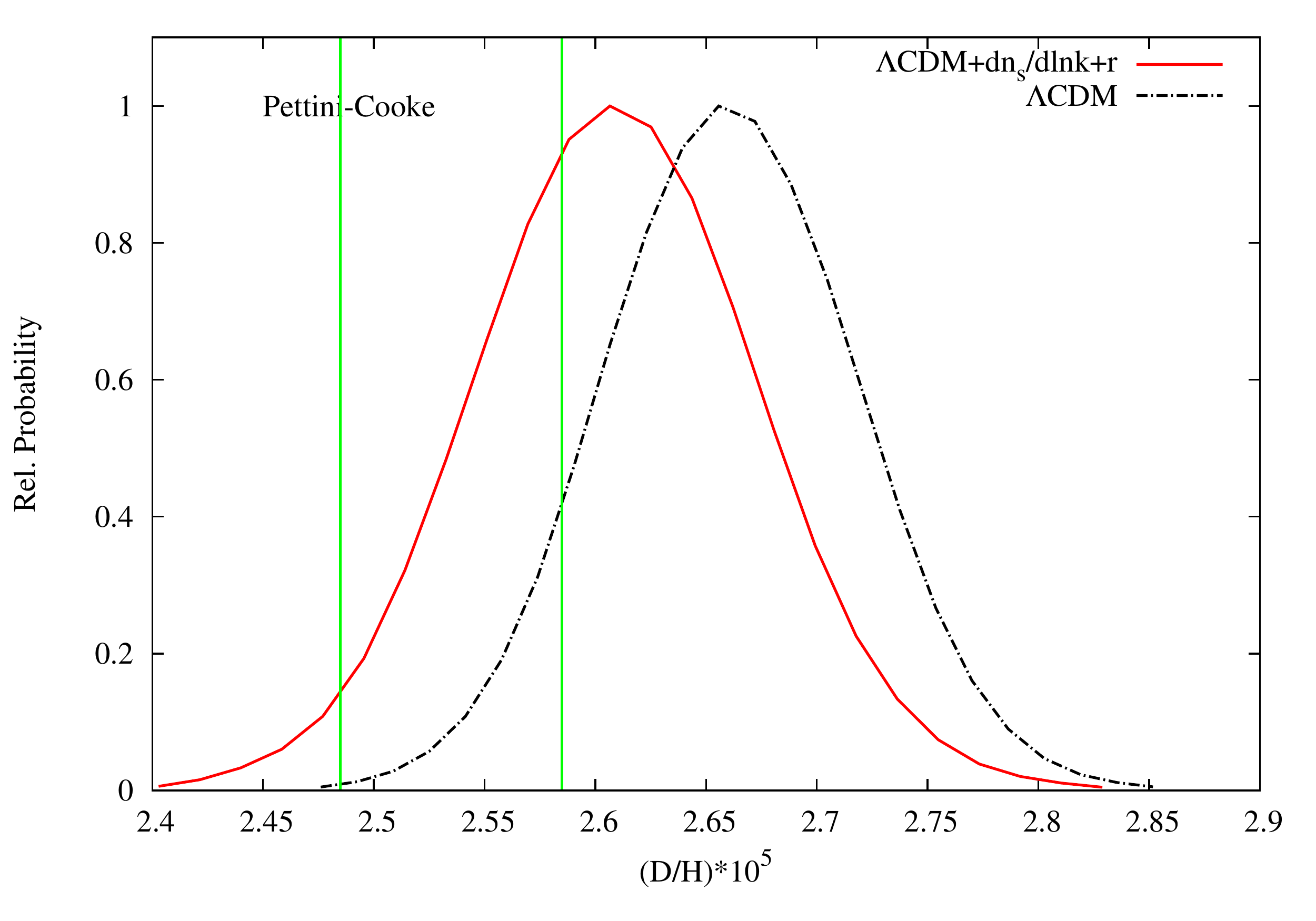} \\ 
 \includegraphics[scale=0.35]{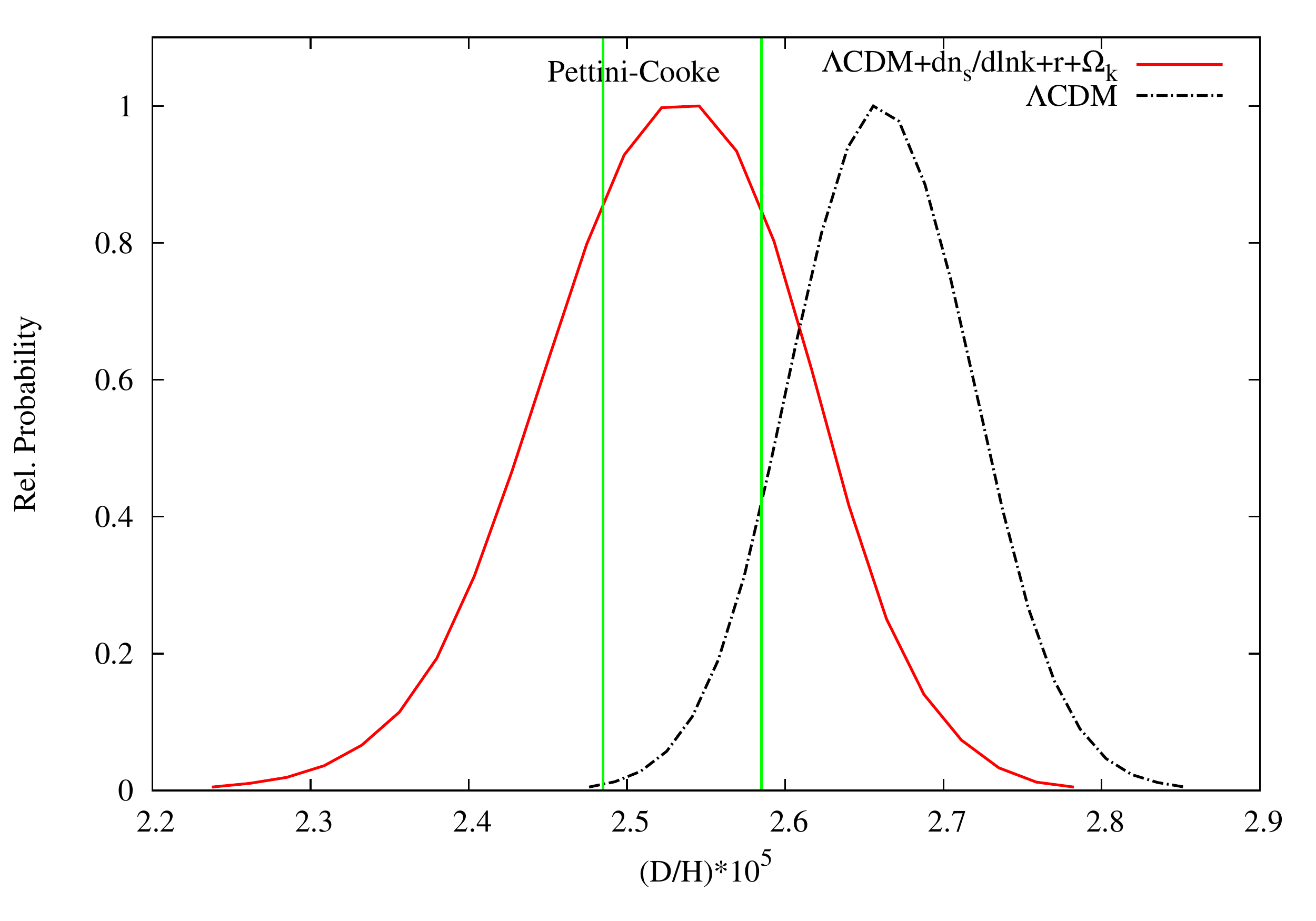}\\%
 \caption{\footnotesize{In this figure we show the posterior distributions for Deuterium in five different models (red line): $\Lambda$CDM$+m_{\nu}+A_{\rm L}$, $\Lambda$CDM$+m_{\nu}+\Omega_{\rm k}$, $\Lambda$CDM$+n_{\nu}+m_{\nu}$, $\Lambda$CDM$+n_{\rm r}+r$ and $\Lambda$CDM$+n_{\rm r}+r+\Omega_{\rm k}$, comparing them with the ones considering the standard model (black line). We report also the $1\sigma$ limits for the results found by Pettini $\&$ Cooke (green lines).}}
 \label{fig:ap3}
\end{figure*}
\end{center}

The posterior distributions of Deuterium are reported, for all the considered models, in Fig.\ref{fig:ap1}-\ref{fig:ap3}.

\begin{center}
 \begin{table}[h!]

TABLE I\\ \vspace{.3cm}
\scalebox{0.8}{
\begin{tabular}{|c|c|c|c|}
\hline 
\rule[-2mm]{0mm}{6mm}
Model & $\Omega_bh^2$ & $(D/H)\cdot 10^{-5}$ & $\dfrac{\Delta}{\sigma_{CMB}}\cdot 10^{-2}$\\
\hline \hline
\rule[-2mm]{0mm}{6mm}
$\Lambda$CDM & $0.02207 \pm 0.00027$ & $2.661\pm0.055$ & $2.3$ \\
\hline
\rule[-2mm]{0mm}{6mm}
+$A_{\rm L}$ & $0.02244 \pm 0.00036$ & $2.585\,\pm 0.067$ & $0.75$\\
\hline
\rule[-2mm]{0mm}{6mm}
+$\alpha_1$ & $0.02216 \pm 0.00029$ & $2.638\,\pm 0.057$ & $1.8$\\
\hline
\rule[-2mm]{0mm}{6mm}
+$\sum m_{\nu}$ & $0.02190 \pm 0.00032$ &  $2.692\,\pm 0.065$ & $2.4$ \\
\hline
\rule[-2mm]{0mm}{6mm}
+$N_{\rm eff}$ & $0.02238 \pm 0.00041$ & $2.753\,\pm 0.094$ & $2.3$\\
\hline
\rule[-2mm]{0mm}{6mm}
+$n_{\rm r}$ & $0.02218 \pm 0.00029$ & $2.636\,\pm 0.057$ & $1.8$\\
\hline
\rule[-2mm]{0mm}{6mm}
+$\Omega_{\rm k}$ & $0.02231 \pm 0.00030$ & $2.611\,\pm 0.059$ & $1.3$ \\
\hline 
\rule[-2mm]{0mm}{6mm}
+$w$ & $0.02206 \pm 0.00028$ & $2.658\,\pm 0.055$ & $2.2$\\
\hline
\rule[-2mm]{0mm}{6mm}
+$r$ & $0.02207 ^{+0.00028}_{-0.00027}$ & $2.656\,\pm 0.055$ & $2.2$\\
\hline
\rule[-2mm]{0mm}{6mm}
+$\sum m_{\nu},\,A_{\rm L}$ & $0.02228 \pm 0.00038$ & $2.617\,\pm 0.074$ & $1.1$\\
\hline
\rule[-2mm]{0mm}{6mm}
+$\sum m_{\nu},\,\Omega_{\rm k}$ & $0.02211 \pm 0.00035$ & $2.650\,\pm 0.068$ & $1.7$ \\
\hline
\rule[-2mm]{0mm}{6mm}
+$\sum m_{\nu},\,N_{\rm eff}$ & $0.02220 \pm 0.00045$ & $2.763\,\pm 0.092$ & $2.5$ \\
\hline
\rule[-2mm]{0mm}{6mm}
+$n_{\rm r},\,r$ & $0.02232 \pm 0.00032$ & $2.611\,\pm 0.061$ & $1.3$ \\
\hline
\rule[-2mm]{0mm}{6mm}
+$n_{\rm r},\,r,\,\Omega_{\rm k}$ & $0.02276 \pm 0.00044 $ & $2.528\,\pm 0.079$ & $0.089$\\
\hline
\end{tabular}}
\caption{\footnotesize{Values of the baryon density $\Omega_bh^2$ and Deuterium for the $\Lambda$CDM model and for all the extensions considered. In the last column is reported the absolute difference with the Pettini $\&$ Cooke value divided by the uncertainty on the CMB estimation.}}
\label{table:1}
\end{table}
\end{center} 

\subsection{Adding the Deuterium Prior} 

To better understand the impact of $D$ on cosmological parameters we use importance sampling technique to include the Pettini $\&$ Cooke $D/H$ measurement to our analysis. In this way chains are weighted imposing a Gaussian prior on $D$. Clearly, we obtain new Deuterium distributions that are more in agreement with the values of Pettini $\&$ Cooke, as expected. However, the interesting part are the shifts in the values of the other parameters.

We find interesting deviations and we report the mean values obtained in Tables \ref{table:2}-\ref{table:6}. 

\begin{center}
\begin{figure*}[h!]
\includegraphics[scale=0.35]{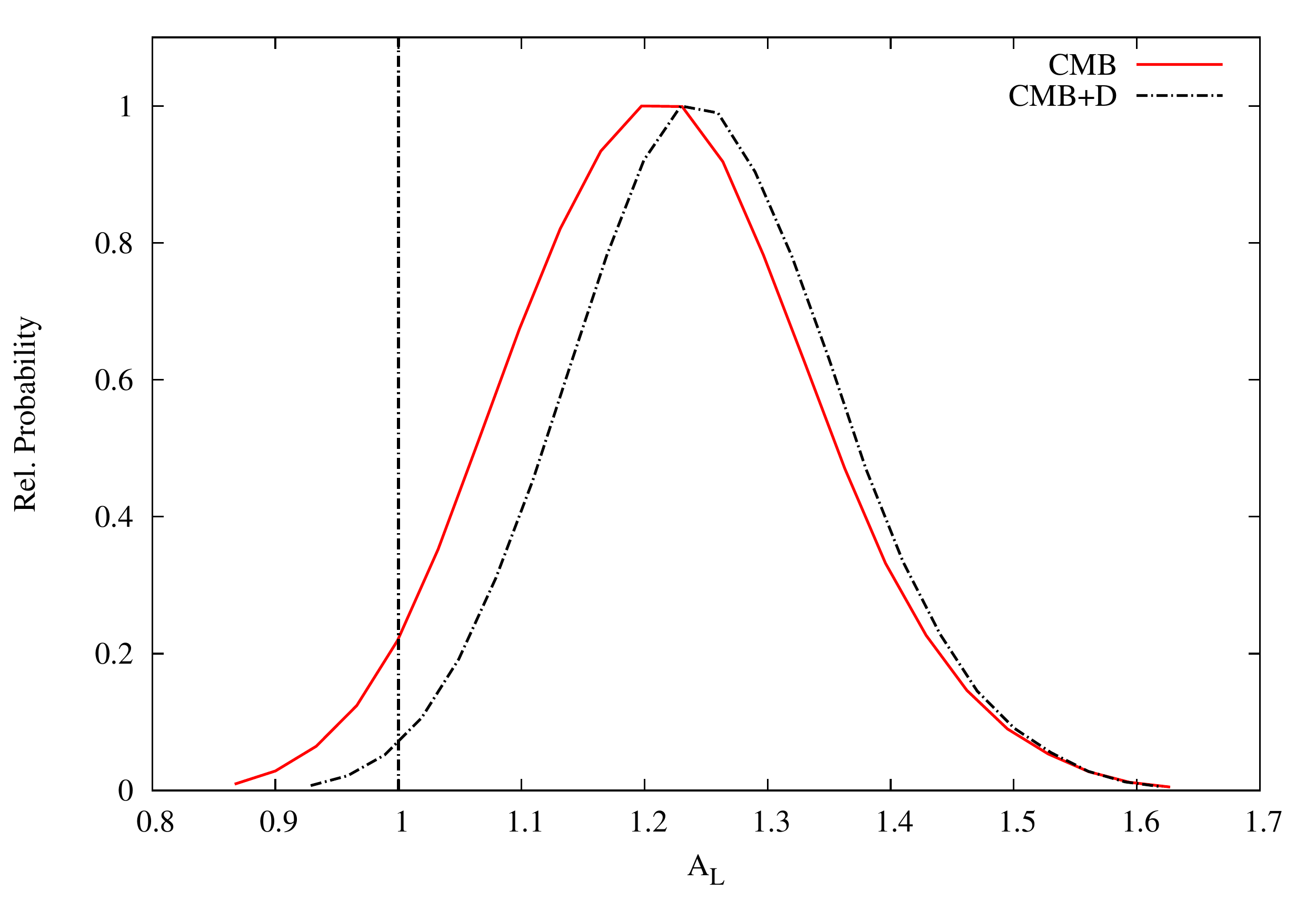}
\includegraphics[scale=0.35]{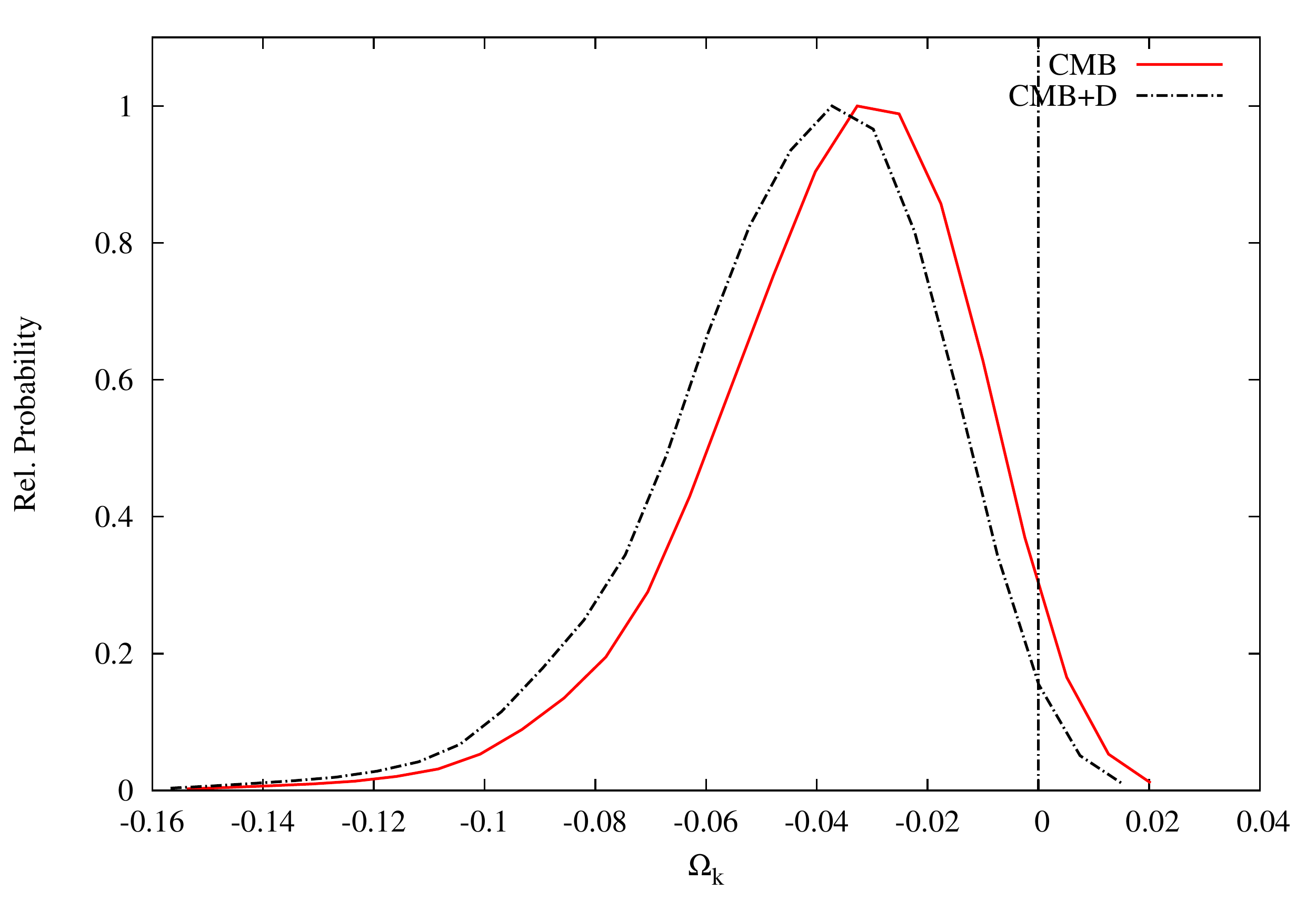}
\caption{\footnotesize{We report here the probability distribution functions for $\Lambda$CDM+$A_{\rm L}$ (left) and $\Lambda$CDM+$\Omega_{\rm k}$ (right) where the red line refers to the CMB-only case, while the black to the results after imposing the $D$ prior.}}
\label{figure:1}
\end{figure*}
\end{center}

\begin{center}
\begin{figure*}[h!]
\includegraphics[scale=0.35]{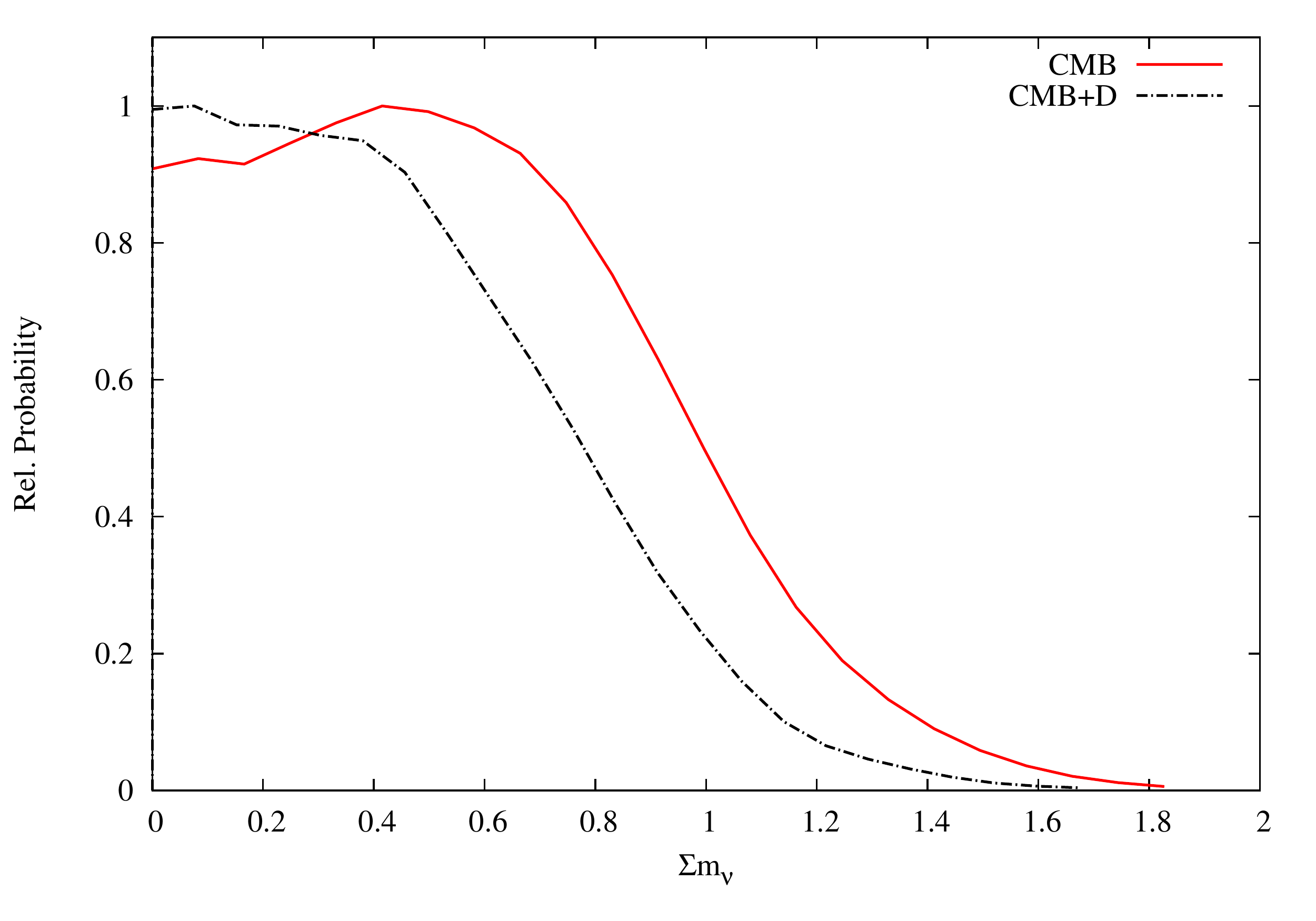}
\includegraphics[scale=0.35]{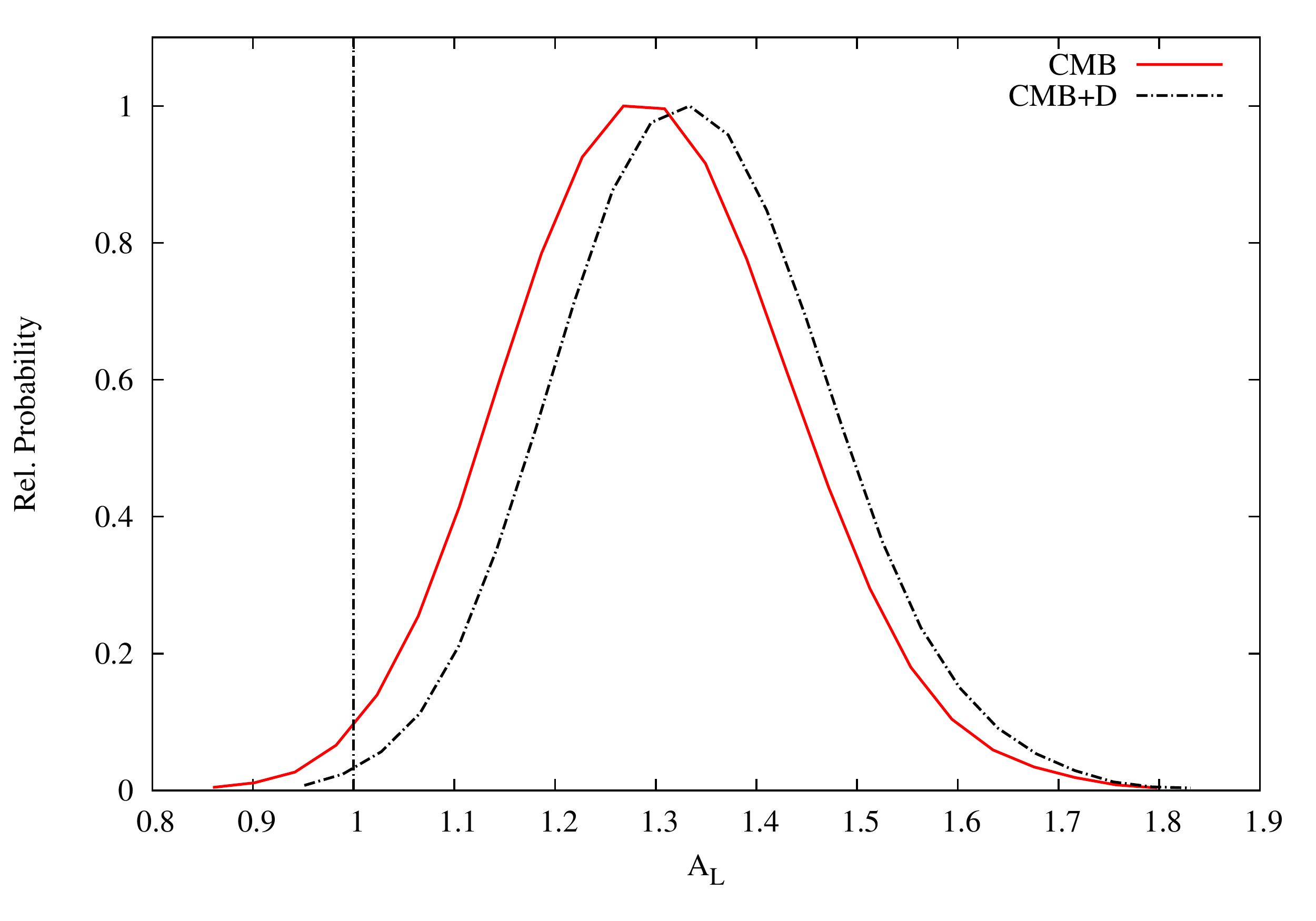}
\caption{\footnotesize{We report here the probability distribution functions for $\Lambda$CDM+$\sum m_{\nu}$+$A_{\rm L}$, where the red line refers to the CMB-only case, while the black to the results after imposing the $D$ prior.}}
\label{figure:2}
\end{figure*}
\end{center}

\begin{center}
\begin{figure*}[h!]
\includegraphics[scale=0.35]{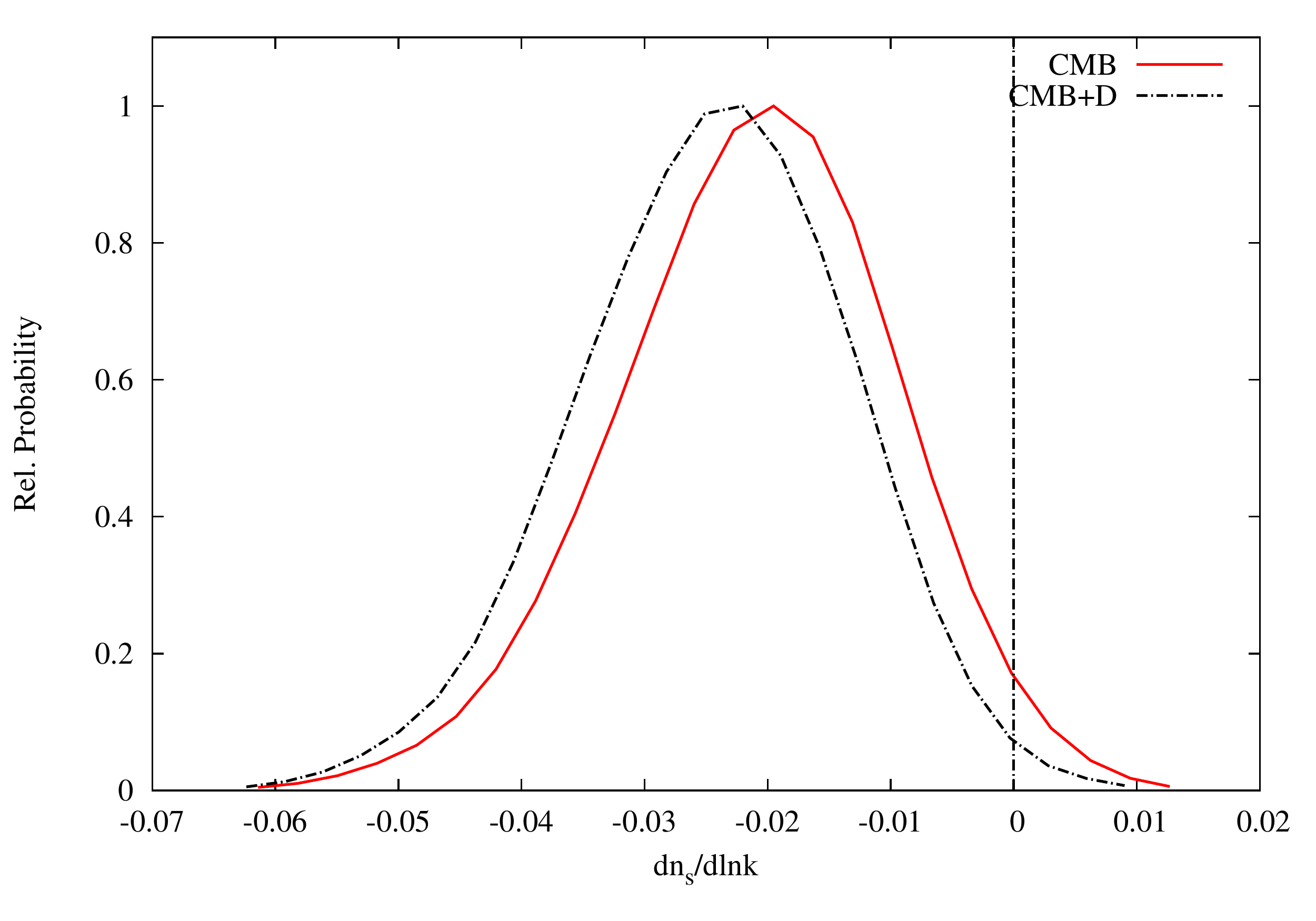}
\includegraphics[scale=0.35]{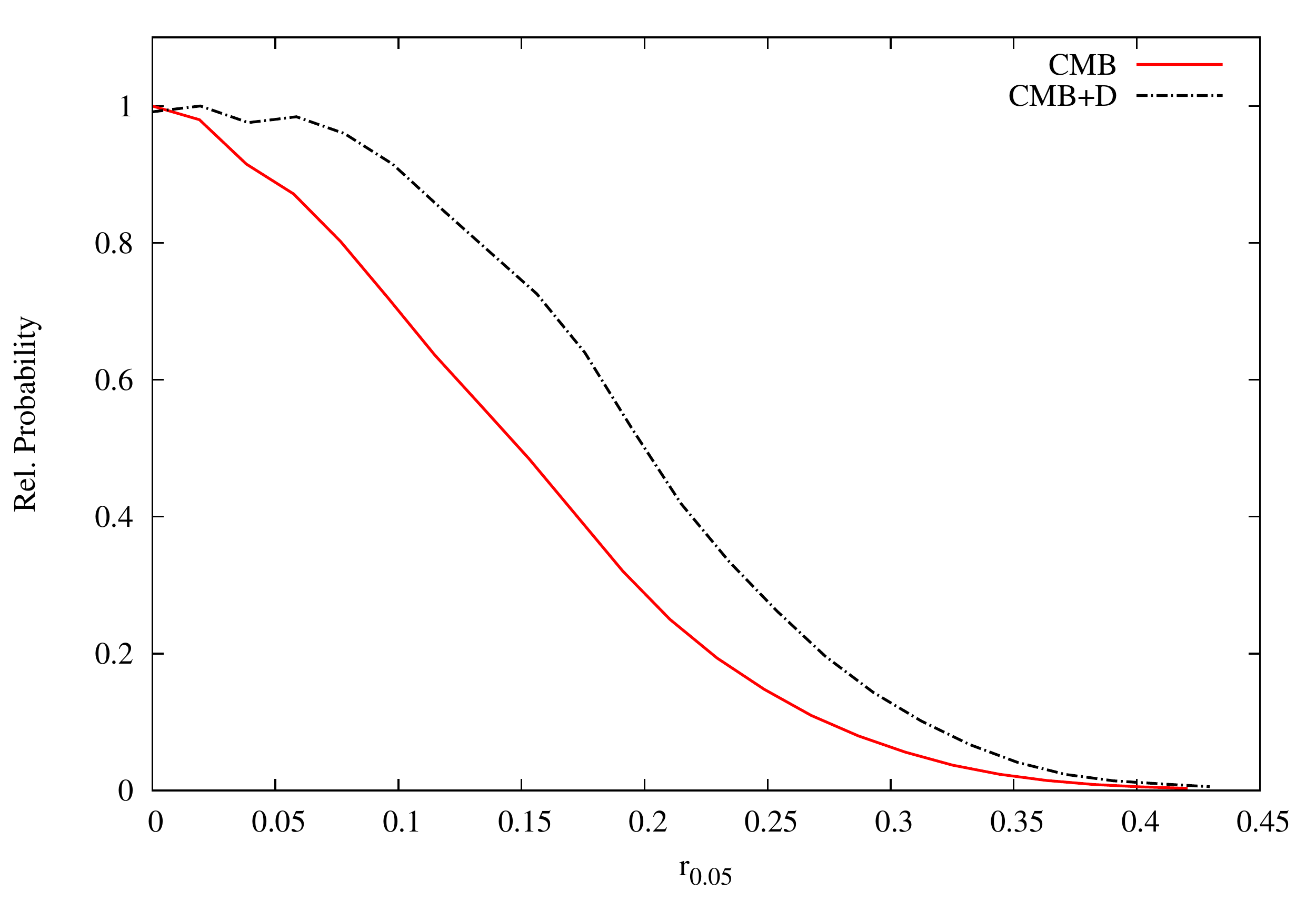}
\caption{\footnotesize{In this figure the probability distribution function for $\Lambda$CDM+$n_{\rm r}$+$r$ is shown, where the red line refers to the CMB-only case, while the black to the results after imposing the $D$ prior.}}
\label{figure:3}
\end{figure*}
\end{center}

\begin{center}
\begin{figure*}[h!]
\includegraphics[scale=0.35]{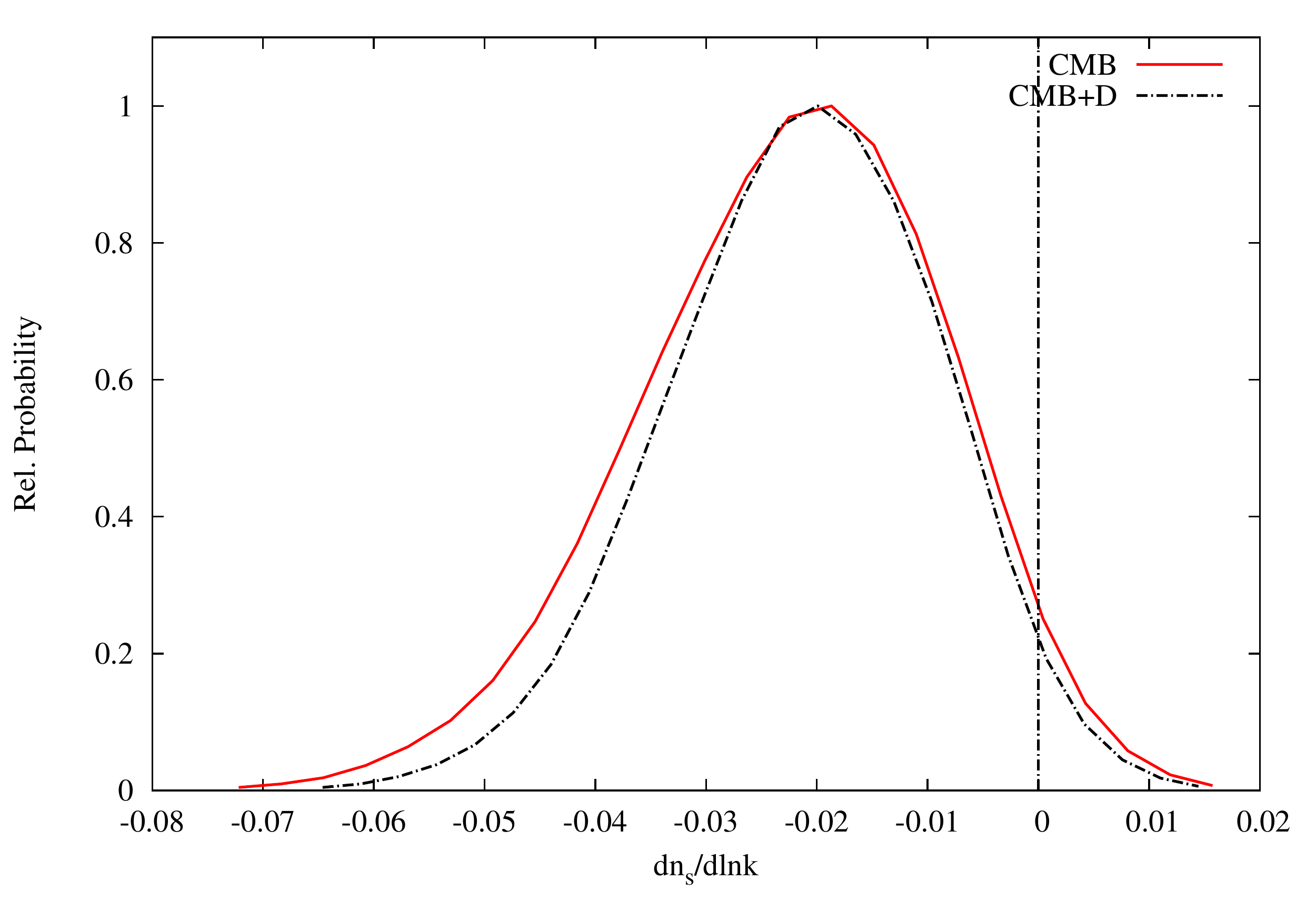}
\includegraphics[scale=0.35]{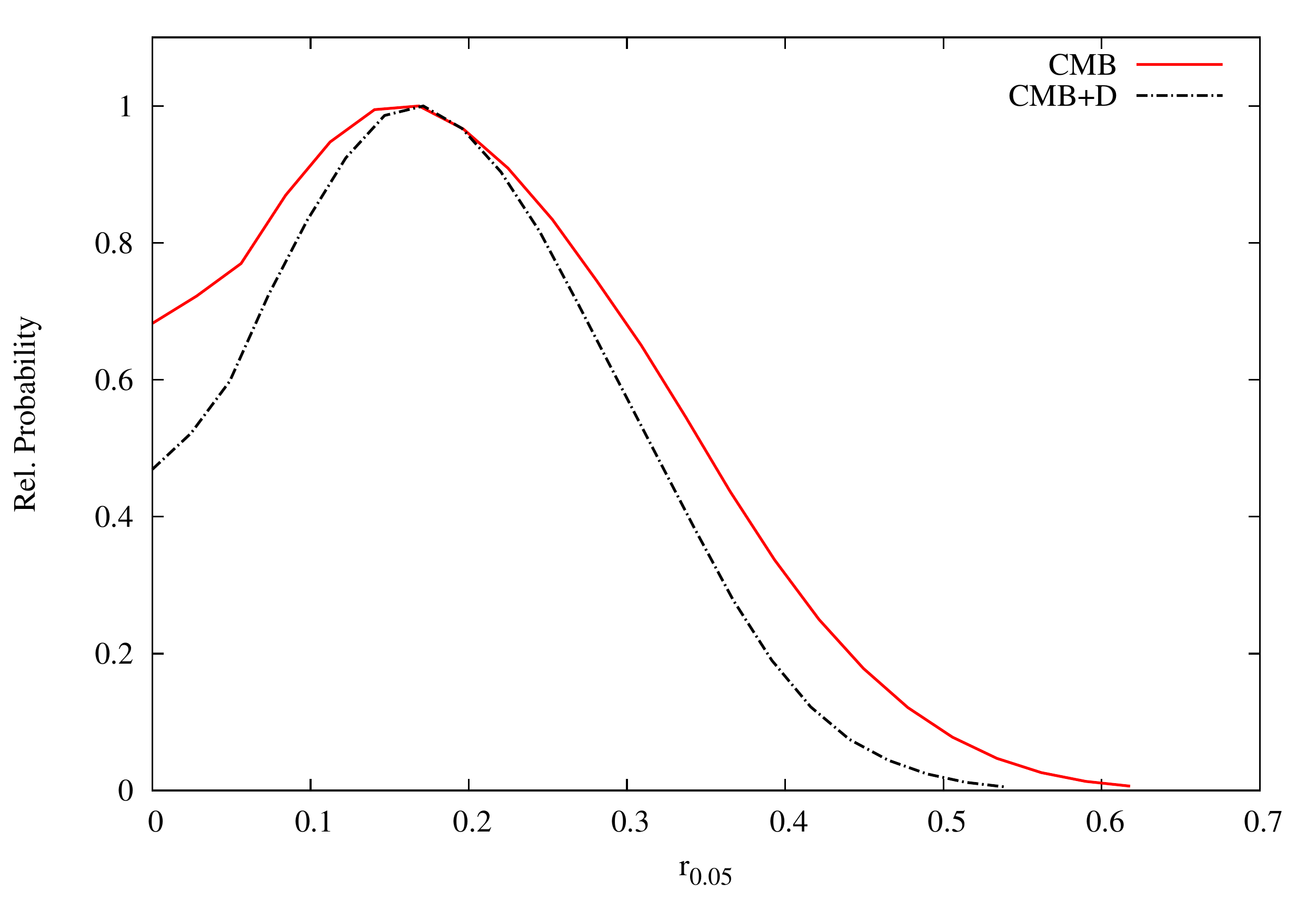}
\includegraphics[scale=0.35]{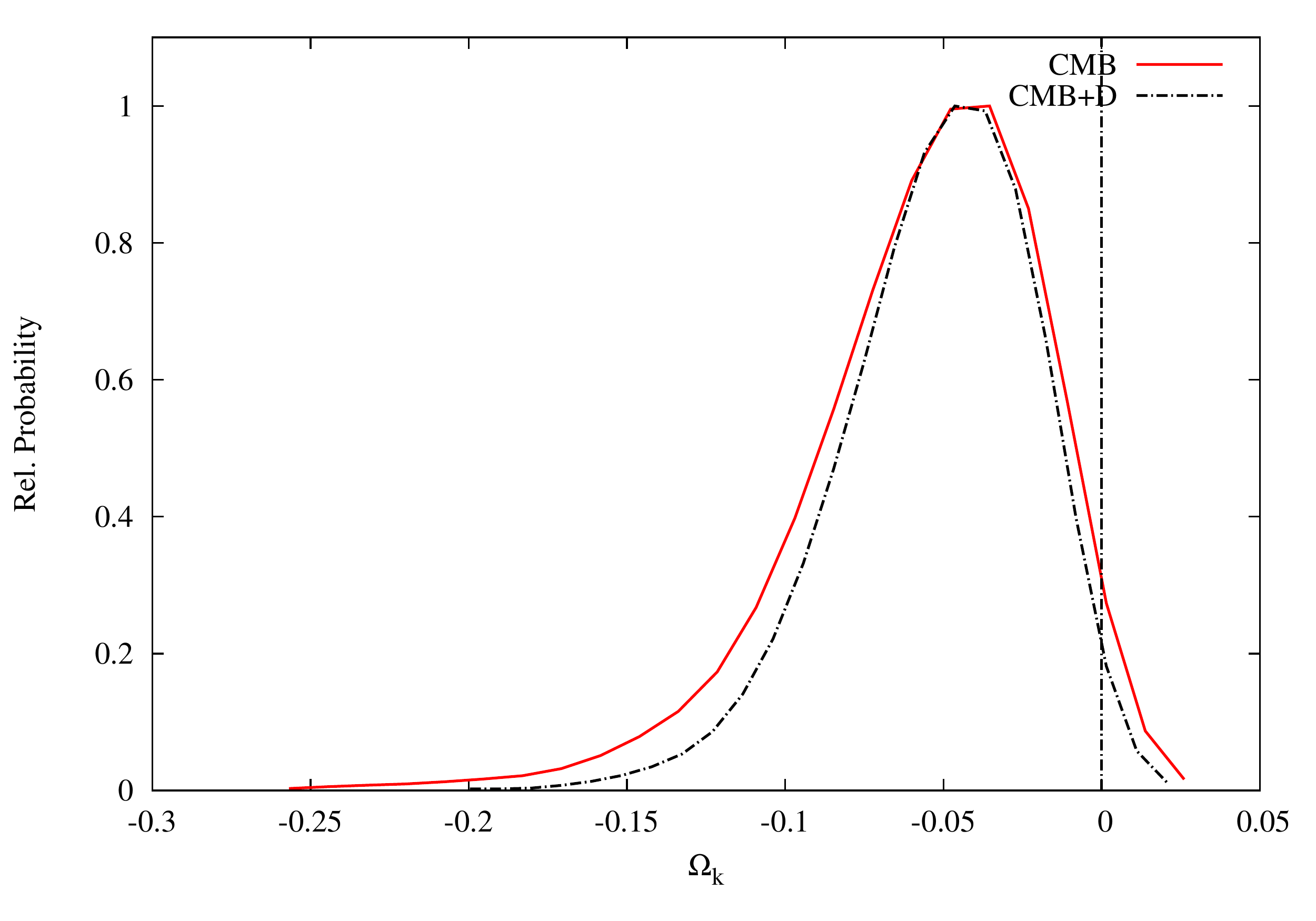}
\caption{\footnotesize{We report in the figure the probability distribution functions for $\Lambda$CDM+$n_{\rm r}$+$r$+$\Omega_{\rm k}$, where the red line refers to the CMB-only case, while the black to the results after imposing the $D$ prior.}}
\label{figure:4}
\end{figure*}
\end{center}

In \ref{table:2} we show the results for the $\Lambda$CDM$+A_{\rm L}$ case: the lensing amplitude parameter is shifted to even higher values with respect to the standard expectation of unity, while the other parameters move towards a general better agreement with WMAP results, showing in particular an higher value for $\Omega_bh^2$. The same effect is seen in the $\Lambda$CDM$+\Omega_{\rm k}$ case, reported in \ref{table:3}: also here we can see that the extra parameter $\Omega_{\rm k}$ tends to depart from the standard value of zero, although pushing the other parameters towards the WMAP values. The  results for the extra parameters are plotted in Fig.\ref{figure:1}. For what concerns the combinations of two parameters, in \ref{table:4} we report the values obtained by simultaneously varying $\sum m_{\nu}$ and $A_{\rm L}$ where we recover again a better agreement with WMAP results for all the parameters but $A_{\rm L}$, which seems to move toward values even higher with respect to the case of \ref{table:2}. The extra parameters posteriors for this case are plotted in Fig.\ref{figure:2}. When $n_{\rm r}$ and $r$ are free to vary we do not see particular deviations for the running of the spectral index, while the tensor to scalar ratio has a slightly higher value (Table \ref{table:5}, Fig.\ref{figure:3}).
Finally we report in Table \ref{table:6} the mean values for the last combination, which includes as free parameters $n_{\rm r}$, $r$ and $\Omega_{\rm k}$. As expected, the introduction of the Deuterium prior does not alter the parameters, given the fact that the CMB estimation in this case is in perfect agreement with the Pettini $\&$ Cooke measurement. The biggest variation is on the value of $r$ that now results different from zero at one $\sigma$. Refer to Fig.\ref{figure:4} for the distributions of extra parameters. 

\section{Conclusions}\label{sec:conclusions}

In this paper we analysed the hint for a tension on the current estimates of the primordial Deuterium abundance obtained indirectly by the Planck CMB measurements and directly by Pettini $\&$ Cooke. Clearly this hint disappears when taking into account the more conservative estimation for the $D$ abundance value from~\cite{Iocco1}. Moreover we also warn the reader that experimental systematics in the
determination of nuclear rates relevant for BBN codes could also be present as showed in \cite{eleo014}.

Assuming that standard BBN is well described by the numerical code \texttt{PArthENoPE},
we obtained different CMB bounds on the primordial deuterium for several theoretical scenarios.

We have found that when two parameters, the lensing amplitude $A_{\rm L}$ and the curvature 
$\Omega_{\rm k}$ are let free to vary a better agreement between the two estimations of $(D/H)$ is obtained. 

A full agreement is found when $n_{\rm r},r$ and $\Omega_{\rm k}$ are simultaneously let free to vary. 
We have found that the variation of other parameters lead to negligible modifications in the CMB
constraints on primordial $D$.

The current tension between the direct measurement of the D abundance and the
indirect, model dependent, CMB bounds can possibly be explained by the introduction of new physics.
It will be the duty of future data, as, for example, the one expected from the Planck 2014 release
that should provide improved constraints on the baryon density, to further confirm the presence of this
tension.

\subsection*{Acknowledgements}
We would like to thank Fabio Iocco for useful advices and support in the making of this paper. We thank also E. Di Valentino, M. Gerbino, A. Marchini and V. Salvatelli for useful discussions.

\clearpage

\begin{center}
\begin{table}[h!]
TABLE II\ \vspace{.3cm}
\scalebox{0.8}{
\begin{tabular}{|c|c|c|}
\hline
\rule[-2mm]{0mm}{6mm}
 & PLANCK+WP & PLANCK+WP+D \\
\hline \hline
\rule[-2mm]{0mm}{6mm}
$A_{\rm L}$ & $1.22\,^{+0.12}_{-0.12}$ & $1.25\,\pm 0.11$ \\
\hline
\rule[-2mm]{0mm}{6mm}
$\Omega_bh^2$ & $(0.2244\,\pm 0.0036)\cdot10^{-1}$ &  $(0.2262\,\pm 0,0022)\cdot10^{-1}$ \\
\hline
\rule[-2mm]{0mm}{6mm}
$\Omega_ch^2$ & $0.1168\,\pm 0.0030$ & $0.1158\,\pm 0.0025$ \\
\hline
\rule[-2mm]{0mm}{6mm}
$n_s$ & $0.9689\,\pm 0.0084$ & $0.9718\,\pm0.0070$ \\
\hline
\rule[-2mm]{0mm}{6mm}
$100\,\theta$ & $1.04181\,\pm 0.00068$ & $1.04201\,\pm0.00061$ \\
\hline
\rule[-2mm]{0mm}{6mm}
$\tau$ & $0.867_{-0.062}^{+0.058}\cdot 10^{-1}$ & $0.877_{-0.063}^{+0.056}\cdot 10^{-1}$ \\
\hline
\rule[-2mm]{0mm}{6mm}
$\text{ln}(10^{10}A_s)$ & $3.077\,\pm 0.025$ & $3.077 \,\pm 0.025$ \\
\hline
\end{tabular}}
\caption{\footnotesize{Parameters mean values and 68\% c. l. for the case $\Lambda$CDM+$A_{\rm L}$. Left column shows results of the CMB-only analysis, while on the right are reported the values obtained after imposing the $D$ prior.}}
\label{table:2}
\end{table}
\end{center}

\begin{center}
\begin{table}[h!]
TABLE III\\ \vspace{.3cm}
\scalebox{0.8}{
\begin{tabular}{|c|c|c|}
\hline
\rule[-2mm]{0mm}{6mm}
 & PLANCK+WP & PLANCK+WP+D \\
\hline \hline
\rule[-2mm]{0mm}{6mm}
$\Omega_{\rm k}$ & $-0.37\,^{+0.23}_{-0.22} \cdot 10^{-1}$ & $-0.44\,\pm0.23 \cdot 10^{-1}$ \\
\hline
\rule[-2mm]{0mm}{6mm}
$\Omega_bh^2$ & $(0.2231\,\pm0.0031)\cdot 10^{-1}$ & $0.2254_{-0.0021}^{+0.0020}\cdot10^{-1}$ \\
\hline
\rule[-2mm]{0mm}{6mm}
$\Omega_ch^2$ & $0.1183_{-0.0029}^{+0.0027}$ & $0.1171\,\pm0.0024$ \\
\hline
\rule[-2mm]{0mm}{6mm}
$n_s$ & $0.9646_{-0.0074}^{+0.0075}$ & $0.9680 _{-0.0067}^{+0.0066}$ \\
\hline
\rule[-2mm]{0mm}{6mm}
$100\,\theta$ & $1.04159 _{-0.00064}^{+0.00065}$ & $1.0418\,\pm0.0060$ \\
\hline
\rule[-2mm]{0mm}{6mm}
$\tau$ & $0.869_{-0.064}^{+0.058}\cdot 10^{-1}$ & $0.892_{-0.064}^{+0.059}\cdot10^{-1}$ \\
\hline
\rule[-2mm]{0mm}{6mm}
$\text{ln}(10^{10}A_s)$ & $3.081\,\pm0.025$  & $3.083\,\pm0.025$ \\
\hline
\end{tabular}}
\caption{\footnotesize{Parameters mean values and 68\% c. l. for the case $\Lambda$CDM+$\Omega_{\rm k}$. Left column shows results of the CMB-only analysis, while on the right are reported the values obtained after imposing the $D$ prior.}}
\label{table:3}
\end{table}
\end{center}

\begin{center}
\begin{table}[h!]
TABLE IV\\ \vspace{.3cm}
\scalebox{0.8}{
\begin{tabular}{|c|c|c|}
\hline 
\rule[-2mm]{0mm}{6mm}
 & PLANCK+WP & PLANCK+WP+D \\
\hline \hline
\rule[-2mm]{0mm}{6mm}
$\sum m_{\nu}$ & $<0.71 \,\text{eV}$  &  $<0.56 \,\text{eV}$ \\
\hline
\rule[-2mm]{0mm}{6mm}
$A_{\rm L}$ & $1.30\,\pm 0.13$   & $1.34\,\pm 0.13$   \\
\hline
\rule[-2mm]{0mm}{6mm}
$\Omega_bh^2$ & $(0.2228\,\pm 0.0038) \cdot 10^{-1}$  & $(0.2257\,\pm 0.0022) \cdot 10^{-1}$   \\
\hline
\rule[-2mm]{0mm}{6mm}
$\Omega_ch^2$ & $0.1177_{-0.0031}^{+0.0030}$ & $0.1160_{-0.0024}^{+0.0025}$  \\
\hline
\rule[-2mm]{0mm}{6mm}
$n_s$ & $0.9644\,\pm 0.0095$   & $0.9699_{-0.0074}^{+0.0073}$  \\
\hline
\rule[-2mm]{0mm}{6mm}
$100\,\theta$ & $1.04144_{-0.00071}^{+0.00072}$ & $1.04178\,\pm 0.00062$   \\
\hline
\rule[-2mm]{0mm}{6mm}
$\tau$ & $0.865_{-0.065}^{+0.058} \cdot 10^{-1}$ & $0.881_{-0.066}^{+0.059} \cdot 10^{-1}$ \\
\hline
\rule[-2mm]{0mm}{6mm}
$\text{ln}(10^{10}A_s)$ & $3.077\,\pm 0.025$  &  $3.077\,\pm 0.026$  \\
\hline
\end{tabular}}
\caption{\footnotesize{Parameters mean values and 68\% c. l. for the case $\Lambda$CDM+$\sum m_{\nu}$+$A_{\rm L}$. Left column shows results of the CMB-only analysis, while on the right are reported the values obtained after imposing the $D$ prior.}}
\label{table:4}
\end{table}
\end{center}

\begin{center}
\begin{table}[h!]
TABLE V\\ \vspace{.3cm}
\scalebox{0.8}{
\begin{tabular}{|c|c|c|}
\hline
\rule[-2mm]{0mm}{6mm}
 & PLANCK+WP & PLANCK+WP+D \\
\hline 	\hline
\rule[-2mm]{0mm}{6mm}
$n_{\rm r}$ & $(-0.21\,\pm 0.11) \cdot 10^{-1}$  & $(-0.24\,\pm 0.11)\cdot 10^{-1}$  \\
\hline
\rule[-2mm]{0mm}{6mm}
$r$ &  $<0.126$  &  $<0.150$ \\
\hline
\rule[-2mm]{0mm}{6mm}
$\Omega_bh^2$ &  $(0.2232\,\pm 0.0032)\cdot 10^{-1}$   &  $(0.2255\,\pm 0.0021)\cdot 10^{-1}$  \\
\hline
\rule[-2mm]{0mm}{6mm}
$\Omega_ch^2$ & $0.1198\,\pm 0.0028$  & $0.1188_{-0.0024}^{+0.0025}$   \\
\hline
\rule[-2mm]{0mm}{6mm}
$n_s$ &  $0.9583\,\pm 0.0080$   &  $0.9610\,\pm 0.0075$  \\
\hline
\rule[-2mm]{0mm}{6mm}
$100\,\theta$ & $1.04140_{-0.00063}^{+0.00062}$  &  $1.04161\,\pm 0.00058$  \\
\hline
\rule[-2mm]{0mm}{6mm}
$\tau$ & $0.1002_{-0.0080}^{+0.0066}$ & $0.1050_{-0.0078}^{+0.0066}$ \\
\hline
\rule[-2mm]{0mm}{6mm}
$\text{ln}(10^{10}A_s)$ & $3.115\,\pm 0.031$   &  $3.123\,\pm 0.031$   \\
\hline
\end{tabular}}
\caption{\footnotesize{Parameters mean values and 68\% c. l. for the case $\Lambda$CDM+$n_{\rm r}$+$r$. Left column shows results of the CMB-only analysis, while on the right are reported the values obtained after imposing the $D$ prior.}}
\label{table:5}
\end{table}
\end{center}

\begin{center}
\begin{table}[h!]
TABLE VI\\ \vspace{.3cm}
\scalebox{0.8}{
\begin{tabular}{|c|c|c|}
\hline
\rule[-2mm]{0mm}{6mm}
 & PLANCK+WP & PLANCK+WP+D \\
\hline \hline
\rule[-2mm]{0mm}{6mm}
$n_{\rm r}$ & $(-0.22\,\pm 0.13)\cdot 10^{-1}$  & $(-0.213\,\pm 0.12)\cdot 10^{-1}$  \\
\hline
\rule[-2mm]{0mm}{6mm}
$r$ &  $<0.253$  &  $0.19\pm0.10$ \\
\hline
\rule[-2mm]{0mm}{6mm}
$\Omega_{\rm k}$ &   $-0.57_{-0.33}^{+0.34}\cdot 10^{-1}$  &  $(-0.52\,\pm 0.29)\cdot 10^{-1}$  \\
\hline
\rule[-2mm]{0mm}{6mm}
$\Omega_bh^2$ &  $(0.2276\,\pm 0.0044)\cdot 10^{-1}$  &  $0.2272_{-0.0024}^{+0.0023} \cdot 10^{-1}$  \\
\hline
\rule[-2mm]{0mm}{6mm}
$\Omega_ch^2$ & $0.1164\,\pm 0.0034$  &  $0.1166\,\pm 0.0027$   \\
\hline
\rule[-2mm]{0mm}{6mm}
$n_s$ &  $0.970\,\pm 0.010$    &  $0.9687_{-0.0083}^{+0.0085}$  \\
\hline
\rule[-2mm]{0mm}{6mm}
$100\,\theta$ & $1.04196_{-0.00073}^{+0.00072}$  &  $1.04191\,\pm 0.00062$  \\
\hline
\rule[-2mm]{0mm}{6mm}
$\tau$ & $0.950_{-0.074}^{+0.066}\cdot 10^{-1}$ & $0.949_{-0.071}^{+0.064}\cdot 10^{-1}$\\
\hline
\rule[-2mm]{0mm}{6mm}
$\text{ln}(10^{10}A_s)$ & $3.097\,\pm -0.030$   & $3.098\,\pm 0.030$   \\
\hline
\end{tabular}}
\caption{\footnotesize{Parameters mean values and 68\% c. l. for the case $\Lambda$CDM+$n_{\rm r}$+$r$+$\Omega_{\rm k}$. Left column shows results of the CMB-only analysis, while on the right are reported the values obtained after imposing the $D$ prior.}}
\label{table:6}
\end{table}
\end{center}


\end{document}